\documentclass[preprint,prd,showpacs,showkeys]{revtex4}
\usepackage{amsmath,graphicx,epsfig,amssymb,amstext}
\allowdisplaybreaks

\def\lh{\textrm{l.h.}}
\def\lQ{\Lambda_{\rm QCD}}
\newcommand{\onec}{1\!\!{\rm l}_c}
\newcommand{\nn}{\nonumber}
\newcommand{\be}{\begin{equation}}
\newcommand{\ee}{\end{equation}}
\newcommand{\bea}{\begin{eqnarray}}
\newcommand{\eea}{\end{eqnarray}}
\def\Dlr{-\frac{i}{2}\overleftrightarrow D}

\def\als{\alpha_{\rm s}}
\def\siml{{\ \lower-1.2pt\vbox{\hbox{\rlap{$<$}\lower6pt\vbox{\hbox{$\sim$}}}}\ }}
\def\simg{{\ \lower-1.2pt\vbox{\hbox{\rlap{$>$}\lower6pt\vbox{\hbox{$\sim$}}}}\ }}

\def\slashchar#1{\setbox0=\hbox{$#1$}\dimen0=\wd0\setbox1=\hbox{/}\dimen1=\wd1\ifdim\dimen0>\dimen1
                 \rlap{\hbox to \dimen0{\hfil/\hfil}} #1 \else \rlap{\hbox to
           \dimen1{\hfil$#1$\hfil}} /  \fi}

\begin{document}
\preprint{IFUM-899-FT}

\title{Hadronic quarkonium decays at order $v^7$}

\author{Nora Brambilla}
\email{nora.brambilla@mi.infn.it}
\author{Antonio Vairo}
\email{antonio.vairo@mi.infn.it}
\affiliation{Dipartimento di Fisica dell'Universit\'a di Milano and INFN, via Celoria 16, 20133 Milano, Italy.}
\author{Emanuele Mereghetti}
\email{emanuele@physics.arizona.edu}
\affiliation{Department of Physics, University of Arizona, Tucson, AZ 85721, USA.}

\begin{abstract}
We compute the complete imaginary part of the NRQCD Lagrangian at order $1/M^4$ 
in the heavy-quark mass expansion, which includes center of mass operators, 
and at order $\als^2$ in the matching coefficients. We also compute 
the imaginary part of the NRQCD Lagrangian at order $1/M^6$ and at order $\als^2$ 
that contributes to the S-wave and P-wave inclusive decay widths of heavy quarkonium into light 
hadrons at order $v^7$ in the heavy-quark velocity expansion.
If we count $\als(M) \sim v^2$, the calculation provides the complete next-to-leading order corrections 
to the P-wave hadronic widths, and in the original NRQCD power counting, the complete next-to-leading order 
corrections to the vector S-wave widths, and part of the next-to-next-to leading order corrections 
to the pseudoscalar S-wave widths. In the S-wave case, we confirm previous findings and add new terms 
in a more conservative power counting. In the P-wave case, our results are in disagreement with previous ones. 
Constraints induced by Poincar\'e invariance on the NRQCD four-fermion sector 
are studied for the first time and provide an additional check of the calculation. 
Perspectives for phenomenological applications are discussed.
\end{abstract}


\maketitle

\section{Introduction}
Non-relativistic effective field theories (NR EFT) of QCD \cite{Brambilla:2004wf,Brambilla:2004jw} 
like non-relativistic QCD (NR\-QCD) \cite{Caswell:1985ui,Bodwin:1994jh} 
offer a systematic framework to access heavy-quarkonium properties and in particular inclusive 
decay widths. Decay width formulas may be organized in a double expansion in the strong coupling constant 
$\als$, calculated at a large scale of the order of the heavy-quark mass $M$, and in the heavy-quark velocity  $v$. 
Both expansion parameters are relatively small. In the bottomonium system, typical reference values 
are $\als(M_b)\approx 0.2$, $v_b^2\approx 0.1$ and in the charmonium one, 
$\als(M_c)\approx 0.35$, $v_c^2\approx 0.3$.

The increasing accuracy of the experimental measurements \cite{Brambilla:2004wf,Amsler:2008zz,Yao:2006px,Eidelman:2004wy} 
calls for a corresponding   accuracy in the theoretical predictions.
The inclusive decay widths of $J/\psi$, $\psi(2S)$ and $\Upsilon(1S)$ into light hadrons are presently known 
within a few percent uncertainty, while the uncertainties in the inclusive decay widths 
of $\Upsilon(2S)$ and $\Upsilon(3S)$ are less than 10\% \cite{Amsler:2008zz}. 
Theoretical accuracies of about 5\% both in the charmonium and in the 
bottomonium case require at least the calculation of ${O}(v^4,\als\,v^2,\als^2)$ corrections.  
The S-wave decay of $\eta_c$ into light hadrons is presently known within a 15\% uncertainty, while the P-wave
decays of $\chi_{c J}$, with $J=0,1,2$, are known within a 10\% uncertainty \cite{Amsler:2008zz}. 
In the P-wave case, the improvement of the experimental accuracy 
has been noticeable over the last few years and the data are now clearly sensitive to 
next-to-leading (NLO) corrections \cite{Vairo:2004sr,Brambilla:2004wf}.
Hence, for the decay of the charmonium P-wave states, theoretical accuracies matching the experimental 
ones require the calculation of ${O}(v^2,\als)$ corrections.  

In this work, we consider relativistic corrections of order $v^2$ and $v^4$ to inclusive decays 
of P- and S-wave quarkonium into light hadrons respectively.
The leading-order S-wave decay width is proportional to the square of the wave-function in the origin
and is therefore of order $v^3$. The leading order P-wave decay width is proportional 
to the square of the derivative of the wave-function in the origin, 
and is therefore of order $v^5$. Then, corrections of order $v^4$ to S-wave decays and 
of order $v^2$ to P-wave decays provide in both cases decay widths at order $v^7$ in the relativistic 
expansion. We consider only processes where the quark and antiquark annihilate into two gluons. 
Hence, more precisely, the paper provides the $\als^2 v^7$ terms of the 
S-wave and P-wave inclusive decay widths. 

In the S-wave case, corrections of order $v^2$ and $v^4$ were first considered 
in \cite{Bodwin:1994jh} and \cite{Bodwin:2002hg} respectively. We agree with their results
if we use their power counting, but find additional contributions in the more conservative 
counting that we adopt. In the P-wave case, corrections of order $v^2$ were first calculated 
in \cite{Huang:1997nt}. Our results disagree with those results. 
In particular, we find different matching coefficients 
for the dimension 10 operators. Moreover, also adopting the power counting of \cite{Huang:1997nt}, 
our decay widths appear to contain two matrix elements more.

The paper is organized in the following way. In section \ref{basics}, we set up
the formalism, discuss the power counting, introduce our basis of operators
and give the general form of the decay widths at order $v^7$.
In section \ref{matching}, we calculate the short-distance imaginary parts 
of the NRQCD four-fermion operators by matching annihilation diagrams of order $\als^2$.
Octet operators are calculated by matching diagrams with an external gluon.
In section \ref{poincare}, we show how Poincar\'e invariance is realized in the EFT 
in the form of exact relations among matching coefficients. Such relations provide 
an additional and independent check of some of the results. In section \ref{conclusions},
we conclude by summarizing the present knowledge about inclusive decays and 
discussing phenomenological applications and future developments of this work.
In appendix \ref{AppA} and \ref{AppB}, we list all the operators and the  matching 
coefficients that have been employed through the paper.

\section{Hadronic decay widths in NRQCD}
\label{basics}
\subsection{NRQCD}
The main mechanism for quarkonium to decay into light hadrons is 
quark-antiquark annihilation. It takes place at a scale which is twice the 
heavy-quark mass $M$. Since this scale is perturbative, quark-antiquark annihilation 
may be described within an expansion in the strong coupling constant $\als$.
Experimentally, this is shown by the narrow widths of quarkonia below the open flavor threshold.
The bound state dynamics, instead, is characterized  by physical scales smaller than 
$M$, such that a perturbative expansion in $\als$ may not be allowed. It is however 
possible to take advantage of the non-relativistic nature of the bound state and expand in the relative 
heavy-quark velocity $v$. In an EFT language, once the scale $M$ has been integrated out, 
the information on the decays is carried by contact terms (four-fermion operators) whose matching 
coefficients develop an imaginary part \cite{Bodwin:1994jh}.
In NRQCD, the decay widths factorize in a high-energy contribution,  encoded
in the imaginary part of the four-fermion matching coefficients,
and a low-energy contribution, encoded in the matrix elements of the
four-fermion operators evaluated on the heavy-quarkonium states.
The NRQCD factorization formula for the inclusive decay width of a quarkonium state $H$ 
into light hadrons (l.h.) is \cite{Bodwin:1994jh}:
\be
\Gamma(H\to {\lh}) = 2 \sum_n
\frac{{\rm Im} \, c^{(n)}}{M^{d_n - 4}}
\langle H |\mathcal O^{(n)}_{\textrm{4-f}} |H \rangle.
\label{factannem}
\ee
$| H\rangle$ is a mass dimension $-{3/2}$ normalized eigenstate
of the NRQCD Hamiltonian with the quantum numbers of the quarkonium state $H$.
The coefficients $c^{(n)}$ can be calculated in perturbation theory by
matching Green functions or physical amplitudes in QCD
and NRQCD. $\mathcal O^{(n)}_{\textrm{4-f}}$ stands for a generic four-fermion operator
of dimension $d_n$, whose general form is $\psi^\dagger (\cdots) \chi \, \chi^\dagger (\cdots) \psi$,
$\psi$ being the Pauli spinor that annihilates a quark and $\chi$ the one that creates an antiquark.
The operators $(\cdots)$ may transform as singlets or octets under color SU(3) gauge transformations.
In the first case, we denote the operator with the subscript $1$, in the second with the subscript $8$.
A list of relevant four-fermion operators is provided in appendix \ref{AppA}.

It is the purpose of this work to calculate the order $\als^2$ contributions to the $c^{(n)}$ coefficients 
that multiply matrix elements up to order $v^7$. These involve operators up to dimension 10.

\subsection{Power Counting}
\label{powcou}
In the factorization formula \eqref{factannem}, the matching coefficients
$c^{(n)}$ are series in $\als$ while the matrix elements
$\langle H |\mathcal O^{(n)}_{\textrm{4-f}} |H \rangle$
are series in $v$ and are, in general, non-perturbative objects.
In NRQCD, several power countings are possible because of the several contributing energy scales. 
These are the relative momentum $Mv$, the binding energy $Mv^2$, and the typical hadronic scale 
$\lQ$; additional scales may enter at higher orders in the calculation \cite{Brambilla:2003mu}.
Whatever power counting one assumes, as long as $v\ll 1$,
matrix elements of operators of higher dimensionality are suppressed by powers of $v$.

The NRQCD Lagrangian is constructed as an expansion in $1/M$ and hence it 
is independent of the power counting. We  shall adopt  a power counting, however, 
when assessing the size of the different matrix elements contributing to the decay widths. 
We will  assume $Mv$ of the same order as $\lQ$ and adopt the following rules.
Matrix elements of the type $\langle H' |\mathcal O |H \rangle$,
where $\mathcal O |H \rangle$ and $|H' \rangle$ have the same quantum numbers 
and color transformation properties in the dominant Fock state,
scale (at leading order) like $(Mv)^{d-3}$, $d$ being the dimension of the operator $\mathcal O$.
If $\mathcal O |H \rangle$ and $|H' \rangle$ do not have the dominant Fock
state with the same quantum numbers, then the matrix element singles out a
component of the quarkonium Fock state that is suppressed.
The amount of suppression depends on the power counting and
on the quantum numbers. As detailed in \cite{Brambilla:2006ph}, the power counting we adopt implies
that the octet components with quantum numbers $S$ and $L\pm 1$, $S$ and $L$,  
$S\pm 1$ and $L$ of a quarkonium state are suppressed by $v$ with respect to the singlet component 
with quantum numbers $S$ and $L$, while the components with $S$, $L\pm 2$ or $S\pm 1$, 
$L \pm 1$ are suppressed by $v^2$.

A different counting, which seems suitable for the situation $Mv^2 \sim \lQ$
has been defined in \cite{Bodwin:1994jh} and used in \cite{Bodwin:2002hg}, \cite{Huang:1997nt}. 
Our power counting is more conservative than the one in  \cite{Bodwin:1994jh}, because 
we assume that all operators scale with the largest available scale, i.e. $Mv \sim \lQ$, 
while in \cite{Bodwin:1994jh} this is not always the case and some operators have extra suppressions.
As a consequence, one may recover the expressions in the power counting of \cite{Bodwin:1994jh} from 
our expressions simply by eliminating matrix elements that, 
in that counting, would be smaller than $v^7$: no new matrix element or matching coefficient 
needs to be added.

For a critical review and a discussion on the different power countings we refer to 
\cite{Brambilla:2004jw} and references therein.

\subsection{Four-fermion operators}
The four-fermion sector of the NRQCD Lagrangian contains 
all four-fermion operators invariant under gauge transformations,  
rotations, translations, charge conjugation, parity and time 
inversion. They may be classified according to their dimensionality and 
color content. The analysis of the four-fermion operators involved in the
hadronic decay widths at order $v^7$ closely parallels the one performed 
for electromagnetic decays in \cite{Brambilla:2006ph}. In the following, we focus on the
main differences, that are mostly related to the contributions of
color octet operators to the hadronic decay widths. The presence
of color octet operators, acting on subleading components of the
heavy-quarkonium Fock state, is one important and well known
characteristics of NRQCD \cite{Bodwin:1994jh}.

We organize the four-fermion sector of the NRQCD
Lagrangian according to the mass dimension and the color structures of the 
operators. In section \ref{redef}, we show how the number of (redundant)
color singlet and octet operators may be reduced by using suitable field redefinitions.
In section \ref{centermass}, we introduce operators proportional to the total momentum 
of the heavy quark-antiquark pair: at variance with the electromagnetic 
case, such operators contribute to the decay widths at order $v^7$.
In appendix \ref{AppA}, we give some details on the construction of octet 
operators of higher dimension and the explicit list of all 
four-fermion operators that need to be considered at the order of accuracy we are working.
Finally, in section \ref{power4f} we use the NRQCD power counting of section \ref{powcou}
to assess the importance of the different matrix elements and 
in section \ref{haddecwidth-general} we write the general form 
of the hadronic decay widths accurate up to order $v^7$.

\subsubsection{Operators from dimension 6 to dimension 10}
\label{dim}
For dimensional reasons, four-fermion operators of mass dimension 6 can only contain
four-fermion fields, without any covariant derivative or gluon field. 
The only allowed color structures are $\onec \otimes \onec$
and $t^a \otimes t^a$. The color octet operator
\begin{equation}\label{eq:4ferm.0}
\psi^{\dag} t^a \chi \chi^{\dag} t^a \psi
\end{equation}
has non vanishing matrix element between the states $\langle (Q\bar Q)_8 g| \ldots |(Q\bar Q)_8 g \rangle$, 
which are subleading components of the heavy-quarkonium Fock state. 
Color octet matrix elements are particularly relevant for
P-wave decays, where they contribute at leading order in the power counting.

Parity conservation forbids four-fermion operators of mass
dimension 7. Four-fermion operators of dimension 8 can be built with two covariant
derivatives or with a chromomagnetic field. For operators built
with two derivatives, the possible color structures are $\onec
\otimes \onec$ and $t^a \otimes t^a$. The construction of color singlet operators 
is straightforward, while some care has to be taken in the color octet case, 
because of the non-Abelian nature of the gauge group, see appendix \ref{AppA}.
The covariant derivatives involved can be proportional either to
the relative momentum of the quark and antiquark pair, for example in an operator like 
\begin{equation}
\label{eq:4ferm.1}
\psi^{\dag} \overleftrightarrow D \chi \cdot \chi^{\dag}
\overleftrightarrow D \psi\,,
\end{equation}
or  to the total momentum of the pair, like in 
\begin{equation*}
\nabla (\psi^{\dag}  \chi) \cdot \nabla (\chi^{\dag}  \psi).
\end{equation*}
Also, operators containing both kind of derivatives can be built, like
\begin{equation*}
   \psi^{\dag}  \left( \Dlr
\right)\times \vec{\sigma} \chi \cdot
          \vec{\nabla} \left( \chi^{\dag}  \psi \right)+\textrm{H.c.}\,.
\end{equation*}

Operators containing the chromomagnetic field can appear with the different color structures 
$t^a \otimes \onec$, $\onec \otimes t^a$, $f^{abc} t^a \otimes t^b$ and $d^{abc} t^a
\otimes t^b$:
\begin{equation}
\label{eq:4ferm.4}
\begin{split}
& \psi^{\dag} g\vec B  \cdot \vec{\sigma}\chi \chi^{\dag} \psi +
\textrm{H.c.}\,,
\\
& \psi^{\dag} g\vec B^a  \cdot \vec{\sigma}\chi \chi^{\dag} t^a
\psi +
\textrm{H.c.}\,,
\\
& f^{abc}\psi^{\dag} g\vec B^a   \cdot \vec{\sigma} t^b\chi
\chi^{\dag} t^c \psi +
\textrm{H.c.}\,,
\\
& d^{abc}\psi^{\dag} g\vec B^a   \cdot \vec{\sigma} t^b\chi
\chi^{\dag} t^c \psi +
\textrm{H.c.}\,.
\\
\end{split}
\end{equation}
Operators of dimension 9 can involve a covariant derivative and a
chromoelectric field,
\begin{equation}\label{eq:4ferm.5}
\psi^{\dag}\chi  \chi^{\dag} (\overleftrightarrow D \cdot
g\vec{E}+ g\vec E \cdot \overleftrightarrow{D} )\psi +
\textrm{H.c.}\,,
\end{equation}
and again we have to consider all the possible color structures,
as in Eq. \eqref{eq:4ferm.4}. Finally, dimension 10 operators may
involve four covariant derivatives or two covariant derivatives
and a chromomagnetic field or two gluon fields. To clarify our
terminology, we call ``singlet operators" the ones in which both
the ingoing and the outgoing $Q\bar Q$ pairs are singlets, as
in \eqref{eq:4ferm.1}, although any covariant derivative also contains an 
octet part, ``octet operators" the ones in which
both the ingoing and the outgoing $Q \bar Q$ pairs are octets, as 
in \eqref{eq:4ferm.0} or in the third and fourth lines of
Eq. \eqref{eq:4ferm.4} and ``singlet-octet transition operators" the
ones in which one of the two pairs is an octet and the other is a 
singlet, as the first two operators of Eq. \eqref{eq:4ferm.4} or the
one in Eq. \eqref{eq:4ferm.5}.
For details on the four-fermion operator definition and construction see appendix \ref{AppA}.

\subsubsection{Field redefinitions}
\label{redef} 
The four-fermion basis built with all possible operators allowed by rotational 
and translational invariance, gauge invariance and invariance under the discrete
symmetries of QCD is redundant since the number of four-fermion operators 
may be reduced by suitable field redefinitions.
The analysis performed in \cite{Brambilla:2006ph} can be extended 
to hadronic singlet operators. Through the field redefinitions 
\be 
\left\{
\begin{gathered}
\psi \rightarrow \psi + \frac{a}{M^5} \left[
\left(\Dlr\right)^2,\chi \chi^\dagger \right]\psi 
\\
\chi \rightarrow \chi - \frac{a}{M^5} \left[
\left(\Dlr\right)^2,\psi \psi^\dagger \right]\chi  
\end{gathered}
\right.  ,
\label{eq:fieldred.1} 
\ee 
it is possible, for a suitable choice of the free parameter $a$, 
to trade the operator $\mathcal T_{1\textrm{-}8}(^1S_0,$ $^1P_1)$, 
defined in Eq. \eqref{eq:B.9}, for the linear combination of $\mathcal Q_1'(^1S_0) - \mathcal
Q_1''(^1S_0) $, defined in Eq. \eqref{eq:B.4}, while, through 
\be
\left\{
\begin{gathered}
\psi \;\;\lower5pt\vbox{\hbox{\rlap{\tiny
$J$}\lower-5pt\vbox{\hbox{$\!
       \rightarrow$}}}}\;\;
\psi + \frac{a}{M^5} \, {\bf T}_{ijlk}^{(J)} \, \sigma^l \left[
\left(\Dlr^i\right)\left(\Dlr^j\right),\chi \chi^\dagger
\right]\sigma^k\psi
\\
\chi \;\;\lower5pt\vbox{\hbox{\rlap{\tiny
$J$}\lower-5pt\vbox{\hbox{$\!
       \rightarrow$}}}}\;\;
\chi - \frac{a}{M^5} \, {\bf T}_{ijlk}^{(J)} \, \sigma^l \left[
\left(\Dlr^j\right)\left(\Dlr^i\right) ,\psi \psi^\dagger
\right]\sigma^k\chi
\end{gathered}
\right., 
\label{eq:fieldred.2} 
\ee 
where 
\bea 
&& {\bf
T}_{ijlk}^{(0)} = \frac{\delta^{ij}\delta^{lk}}{3}, \label{t0}
\\
&& {\bf T}_{ijlk}^{(1)} = \frac{\epsilon_{ijn}\epsilon_{kln}}{2}, \label{t1}
\\
&& {\bf T}_{ijlk}^{(2)} = \frac{\delta^{il}\delta^{jk} +
\delta^{jl}\delta^{ik}}{2} -\frac{\delta^{ij}\delta^{lk}}{3}, \label{t2} 
\eea
the operators $\mathcal T^{(i)}_{1\textrm{-}8}(^3S_1,^3P)$, with $i = 0,1,2$, can be eliminated 
by a suitable choice of $a$ and by redefining the matching coefficients of $\mathcal Q'_1(^3S_1)$,
$\mathcal Q_1''(^3S_1)$, $\mathcal Q_1'(^3S_1, ^3D_1)$ and
$\mathcal Q''_1(^3S_1,^3D_1)$ (see Eqs. \eqref{eq:B.9} and
\eqref{eq:B.4} for the definition of these operators). As it was
noted in \cite{Brambilla:2006ph}, these field redefinitions do not
change the sums of the coefficients $h'_1(^1S_0)+h''_1(^1S_0)$,
$h'_1(^3S_1)+h''_1(^3S_1)$ and $h'_1(^3S_1,^3D_1)+h''_1(^3S_1,^3D_1)$.

It is also possible to exploit field redefinitions to reduce the number of 
octet operators. Consider the field redefinitions
\begin{equation}
\left\{
\begin{gathered}
\psi \rightarrow \psi + \frac{a}{M^5} \left[
\left(-\frac{i}{2}\overleftrightarrow D\right)^2, t^a\chi
\chi^\dagger \right] t^a\psi 
\\
\chi \rightarrow \chi - \frac{a}{M^5} \left[
\left(-\frac{i}{2}\overleftrightarrow D\right)^2,  t^a \psi
\psi^\dagger \right] t^a \chi
\end{gathered}
\right., 
\label{eq:fieldred.3}
\end{equation}
where the definition of $\psi^{\dag} \overleftrightarrow D^2 t^a
\chi$ is given in Eq. \eqref{eq:gauge.6}. Eq. \eqref{eq:fieldred.3}
induces the following transformation
\bea
\psi^{\dag} i D_0 \psi + \chi^{\dag} i D_0 \chi &\rightarrow&
\psi^{\dag} i D_0 \psi + \chi^{\dag} i D_0 \chi - \frac{a}{M^5}
\frac{1}{N_c} \mathcal T_{1\textrm{-}8}(^1P_1,^1S_0) 
\nn\\
&&
- \frac{a}{2M^5}\mathcal D_{8\textrm{-}8}(^1S_0,^1P_1)
+ \frac{a}{2M^5}\mathcal F_{8}(^1S_0),
\label{eq:fieldred.4}
\eea
where $N_c =3$ is the number of colors and the operators $\mathcal D_{8\textrm{-}8}(^1S_0,^1P_1)$ 
and $\mathcal F_{8}(^1S_0)$ are defined in Eq. \eqref{eq:B.9}.
The same field redefinitions induce the following transformation on the kinetic term
\begin{equation}
\label{eq:fieldred.8}
\begin{split}
\psi^{\dag}   \frac{\vec D^2}{2M}  \psi - \chi^{\dag}  \frac{\vec
D^2}{2M}  \chi \rightarrow & \,\,\psi^{\dag}  \frac{\vec D^2}{2M}
\psi - \chi^{\dag} \frac{\vec D^2}{2M} \chi \\ &+ \frac{2a}{M^6}
\left( \mathcal Q'_8(^1S_0) - \mathcal Q''_8(^1S_0) \right) ,
\end{split}
\end{equation}
where in the right-hand side we have neglected operators proportional to
the center of mass momentum of the quark-antiquark pair. Equations \eqref{eq:fieldred.4} and
\eqref{eq:fieldred.8} show that the operators $\mathcal T_{1\textrm{-}8}(^1P_1,^1S_0)$  and 
$\mathcal Q'_8(^1S_0)- \mathcal Q''_8(^1S_0)$ are not independent
and that it is possible, for a suitable choice of the parameter $a$,  
to trade the one for a redefinition of the matching coefficient
of the other and of $\mathcal D_{8\textrm{-}8}(^1S_0,^1P_1)$ and $\mathcal F_{8}(^1S_0)$.

With a closely related argument, introducing the field redefinitions
\begin{equation}
\left\{
\begin{gathered}
\psi 
\;\;\lower5pt\vbox{\hbox{\rlap{\tiny
$J$}\lower-5pt\vbox{\hbox{$\!
       \rightarrow$}}}}\;\;
\psi + \frac{a}{M^5} {\bf T}_{ijlk}^{(J)}
\sigma^l \left[ \left(-\frac{1}{4}\overleftrightarrow
D^i\overleftrightarrow D^j\right),  t^a\chi \chi^\dagger \right]
t^a\sigma^k\psi 
\\
\chi 
\;\;\lower5pt\vbox{\hbox{\rlap{\tiny
$J$}\lower-5pt\vbox{\hbox{$\!
       \rightarrow$}}}}\;\;
\chi - \frac{a}{M^5} {\bf T}_{ijlk}^{(J)}
\sigma^l \left[ \left(-\frac{1}{4}\overleftrightarrow
D^j\overleftrightarrow D^i \right), t^a \psi \psi^\dagger \right] 
t^a \sigma^k \chi
\end{gathered}
\right., 
\label{eq:fieldred.9}
\end{equation}
with ${\bf T}_{ijlk}^{(J)}$ given in Eqs. \eqref{t0}-\eqref{t2} 
and $\overleftrightarrow D^i \overleftrightarrow D^j t^a$  
defined according to Eq. \eqref{eq:gauge.6}, it is possible to set the parameter $a$ 
in such a way that the minimal basis of operators either contains the three operators 
$\mathcal T_{1\textrm{-}8}(^3P_J,^3S_1)$ or, with a different choice of $a$, 
the three operators  
$1/2( \mathcal Q'_8(^3S_1)  - $ $ \mathcal Q''_8(^3S_1) )$, 
$1/2 (\mathcal Q'_8(^3S_1,^3D_1)$ $-$ $ \mathcal Q''_8(^3S_1,^3D_1) )$ and 
$\mathcal T_{8\textrm{-}1}^{(1)\prime}(^3S_1,^3P)$ defined in Eqs. \eqref{eq:B.9}
and \eqref{eq:B.4}. The first set of operators is more useful in dealing with P-wave decay 
widths and we will use it in the rest of the paper.

Note that the operator $\mathcal T_{8\textrm{-}1}^{(1)\prime}(^3S_1,^3P)$ as well as the 
operators  $\mathcal T_{1\textrm{-}8}^{(i)}(^3S_1,^3P)$ previously introduced 
and $\mathcal T_{1\textrm{-}8}^{(1)\prime}(^3S_1,^3P)$, which is required 
by the matching, annihilate (create) a singlet $Q \bar Q$ pair 
with orbital angular momentum $L=1$ but with no definite value of $J$. 
So, in our notation, we denote the annihilated pair just with its spin and orbital angular 
momentum quantum numbers, omitting the subscript $J$.

\subsubsection{Operators proportional to the total momentum of the quark-anti\-qu\-ark pair}
\label{centermass} 
The description of the hadronic decay widths up to order $v^7$ requires 
the inclusion of operators proportional to the total momentum of the
quark-antiquark pair into the meson. By parity conservation these operators 
must contain at least two derivatives,  so they have at least mass dimension 8.
The two derivatives can act on the $Q \bar Q$ pair, like in
\begin{equation}
\label{eq:com.1}
\mathcal P_{1a\, \textrm{cm}}  = \vec{\nabla}^i \left(\psi^{\dag}
\sigma^j \chi\right)
 \vec{\nabla}^i (\chi^{\dag} \sigma^j \psi) .
\end{equation}
Since the $Q \bar Q$ pair is a color singlet, $\vec{\nabla}$ is an ordinary derivative.
If the $Q \bar Q$ pair is a color octet, we can build an operator analogous to \eqref{eq:com.1}
\begin{equation}
\label{eq:com.2}
\mathcal P_{8a\, \textrm{cm}}  = \vec{D}^i_{ab}  \left(\psi^{\dag}
t^b \sigma^j \chi\right) \vec{D}^i_{ac}  \left( \chi^{\dag}
t^c \sigma^j \psi\right),
\end{equation}
where ${\vec D}_{ab}$ is a covariant derivative in the adjoint representation.

Also operators containing a total derivative $\vec\nabla$ and a 
derivative $\overleftrightarrow D$, proportional to the relative momentum of the pair,
can be built. In this case, since under charge conjugation 
${\vec \nabla}(\psi^{\dag} \chi) \rightarrow {\vec \nabla}(\psi^{\dag} \chi)$ 
and $\psi^{\dag} \overleftrightarrow D \chi \rightarrow - \psi^{\dag}
\overleftrightarrow D \chi$, the operators must contain a  Pauli matrix in order 
to be charge conjugation invariant. An example is the operator
\begin{equation}
\label{eq:com.3}
\mathcal O_{1\,\textrm{cm}}  =   \psi^{\dag}  \left( \Dlr
\right)\times \vec{\sigma} \chi \cdot
          \vec{\nabla} \left( \chi^{\dag}  \psi \right) +\textrm{H.c.}\,.
\end{equation}
As explained in section \ref{poincare}, the matching coefficients of the operators 
of mass dimension 8 proportional to the total momentum of the $Q \bar Q$ 
pair are completely determined by the coefficients of the dimension 6 operators.
These relations are a manifestation of the Poincar\'{e} invariance of the effective field theory.

\subsubsection{Power counting of the four-fermion operators}
\label{power4f}
From the rules given in  section \ref{powcou}, it follows that 
\be
\langle H(^{2S+1}L_J) | \frac{1}{M^{d-4}} \, {\mathcal O}_1(^{2S+1}L_J)| H(^{2S+1}L_J) \rangle
\sim M v^{d-3},
\ee
where $| H(^{2S+1}L_J) \rangle$ stands for a quarkonium state whose
dominant Fock-space component is a $Q \bar Q$ pair with quantum numbers $S$,
$L$ and $J$, ${\mathcal O}_1(^{2S+1}L_J)$ is a singlet 
four-fermion operator that acts on the  $Q \bar Q$ pair with  spin $S$, 
orbital angular momentum $L$ and total angular momentum $J$ and $d$ 
is the dimension of the operator.

The scaling of color octet matrix elements is affected by the suppression
of the Fock state component they act on. For example, the power counting given in section \ref{powcou} implies
\begin{equation}
\label{countoct}
\begin{split}
\langle H(^3P_0) | \frac{1}{M^{2}} \, {\mathcal O}_8(^3S_1)| H(^3P_0) \rangle & \sim M v^{5}, 
\\
\langle H(^1S_0) | \frac{1}{M^{2}} \, {\mathcal O}_8(^3S_1)| H(^1S_0) \rangle & \sim M v^{5}.
\end{split}
\end{equation}

In the power counting that we adopt, the gluon field and the derivative that 
belong to a covariant derivative have the same scaling. If the gluon 
field selects a component of the quarkonium Fock state, which is suppressed, like in 
$\langle H(^3P_0) | {\mathcal O}_1(^3P_0)| H(^3P_0) \rangle$, then its contribution 
to the matrix element is subleading. If, however, the gluon field  
selects a component whose projection on the operator is not suppressed 
or the gluon is reabsorbed by other gluons in the operator, then it may happen 
that the gluon part in the covariant derivative gives to the matrix element 
a contribution that is larger than the one provided by the derivative part.
For example, due to the gluons in the covariant derivatives, dimension 10 octet operators 
like  $\mathcal P_8(^1P_1)$, $\mathcal Q'_8(^1S_0)$ and $ \mathcal Q_8(^1D_2)$, 
as well as the singlet operator $ \mathcal Q_1(^1D_2)$, contribute 
to the decay width of the quarkonium state $H(^1S_0)$ at order $v^7$.
Similar operators contribute at order $v^7$ also to the decay width of  the quarkonium states 
$H(^3S_1)$ and  $H(^3P_J)$. 

Concerning the scaling of the singlet-octet matrix elements, in the power counting of  section \ref{powcou}
both the chromoelectric and chromomagnetic fields scale as their mass dimension, $(Mv)^2$, 
so the scaling of a matrix element is $M v^{d-3} v^s$, where $v^s$ takes into account the suppression
of the Fock state the operator acts on. 
For example, consider the matrix elements of the dimension 8 operators defined 
in Eq. \eqref{eq:B.8}:
\be
\langle H(^1S_0) |  \frac{1}{M^{4}}\, \mathcal S_{1\textrm{-}8}(^1S_0,^3S_1) | H(^1S_0) \rangle,
\ee
and 
\be
\langle H(^3S_1) |  \frac{1}{M^{4}}\, \mathcal S_{1\textrm{-}8}(^3S_1,^1S_0) | H(^3S_1) \rangle.
\ee
The operator $\mathcal S_{1\textrm-8}(^1S_0,^3S_1)$ 
destroys a singlet $Q \bar Q$ pair with quantum numbers $^1S_0$ and creates
an octet $Q \bar Q$ pair with quantum numbers $^3S_1$ and a gluon (and viceversa),
the operator $\mathcal S_{1\textrm-8}(^3S_1,^1S_0)$ 
destroys a singlet $Q \bar Q$ pair with quantum numbers $^3S_1$ and creates
an octet $Q \bar Q$ pair with quantum numbers $^1S_0$ and a gluon (and viceversa).
Hence, both matrix elements scale like $M v^6$.

Equations \eqref{eq:B.9} define octet operators of dimension 9, and
since the octet Fock-space component is suppressed by $v$, we have 
\be
\langle H(^3S_1) |  \frac{1}{M^{5}}\, \mathcal T^{(1) \prime}_{1\textrm{-}8}(^3S_1,^3P)  | H(^3S_1) \rangle 
\sim M v^7,
\ee
and 
\be 
\langle H(^3P_J) |  \frac{1}{M^{5}}\, \mathcal T_{1\textrm{-}8}(^3P_J,^3S_1)  | H(^3P_J) \rangle \sim M v^7.
\ee
For the reasons discussed above, in our power counting, matrix elements of octet operators of dimension 9, 
like $\mathcal D_{8\textrm{-}8}(^1S_0,^1P_1)$, are not necessarily negligible at order $v^7$ 
because of the gluons in the covariant derivatives, which may couple to other gluons 
in the operator and in the quarkonium Fock state. For instance, we have 
\begin{equation}
\label{octoct}
\langle H(^1S_0) | \frac{1}{M^5} \mathcal D_{8\textrm{-}8}(^1S_0,^1P_1) | H(^1S_0) \rangle \sim M v^7.
\end{equation}
Matrix elements of the operator $\mathcal F_{8}(^1S_0)$ are smaller than $v^7$ because of the 
suppression induced by the Gauss law.
Note that also the matrix element of the following dimension 10 operator is negligible at order $v^7$:
\be
\langle H(^3P_0) | \frac{1}{M^{6}}\, \psi^{\dag}  \vec B  \cdot \overleftrightarrow D \chi  
\chi^{\dag} \overleftrightarrow D \cdot \vec{\sigma} \psi | H(^3P_0) \rangle  \sim M v^8.
\ee

Finally, we discuss the scaling of matrix elements of operators 
proportional to the total momentum of the $Q \bar Q$ pair.
We work in a frame in which the heavy quarkonium is at rest. In
this frame, operators proportional to the total momentum of the
pair have non vanishing matrix elements only between subleading
components of the heavy-quarkonium Fock state, containing at least
one gluon. Lattice data indicate that higher gluonic excitations between 
the $Q \bar Q$ pair are separated from the lowest quarkonium state by a mass 
gap of oder $\lQ$ (for a detailed discussion, see \cite{Brambilla:2004jw} 
and references therein). Therefore, gluons in subleading components 
of the Fock space must be counted as soft $(q^0,\vec q \,) \sim (Mv, Mv)$, where $Mv  \sim \lQ$. 
The emission of a soft gluon leaves the $Q \bar Q$ pair with a
total momentum of order $Mv$, hence, the scaling of the operators
$\vec{\nabla}$ and $\vec D_{ab}$ acting on the $Q \bar Q$ pair is $ \sim Mv$.
Consider, for example, the matrix element of the operator $\mathcal O_{8 \, \textrm{cm}}$  between $^3S_1$ states
\begin{equation}
\label{eq:pow4f.1}
\langle H(^3S_1) | \mathcal O_{8\,\textrm{cm}}| H(^3S_1) \rangle =
\langle ^3S_1 | \psi^{\dag} t^a \left( \Dlr \right)\times
\vec{\sigma} \chi \cdot
          \vec{D}_{ab} \left( \chi^{\dag} t^b \psi \right) 
+ \textrm{H.c.}
| (^1S_0)_8 g \rangle + \dots \, .
\end{equation}
The leading order contribution to the l.h.s. of Eq. \eqref{eq:pow4f.1} comes from the matrix element 
between the components $| ^3S_1 \rangle$ and  $| (^1S_0)_8 g\rangle$ of $|H(^3S_1)\rangle$, 
the gluon in the incoming state being annihilated by the gluon field in $\overleftrightarrow D$. The
matrix element in Eq. \eqref{eq:pow4f.1} gets a $v$ suppression from
each derivative, and a further $v$ suppression from the $|(^1S_0)_8 g \rangle$ state. Therefore it 
scales like $v^6$ and  is suppressed by $v^3$ with respect to the leading 
contribution to the decay width. The operator $\mathcal P_{8 a \, \textrm{cm}}$ has  nonvanishing matrix
element if both the incoming and outgoing states contain a gluon.
For example, it contributes to the decay width of $H(^3P_0)$:
\begin{equation}
\label{eq:pow4f.2}
\langle H(^3P_0) |\mathcal P_{8 a \,\textrm{cm}} |H(^3P_0) \rangle  =
\langle (^3S_1)_8 g |\vec{D}^i_{ab}  \left(\psi^{\dag} t^b
\sigma^j \chi\right) \vec{D}^i_{ac} \cdot \left( \chi^{\dag} t^c
\sigma^j \psi\right) 
+ \textrm{H.c.}
| (^3S_1)_8 g\rangle + \dots \, .
\end{equation}
The matrix element in Eq. \eqref{eq:pow4f.2} gets two powers of $v$
from the derivatives and two from the states, so it scales like
$v^7$, and contributes to the P-wave decay width at the order we are interested in. 

We note that for electromagnetic decays, operators proportional 
to the total momentum of the $Q \bar Q$ pair do not contribute to decay 
widths calculated in the quarkonium center of mass rest frame. 
The reason is the following. Electromagnetic operators are obtained by inserting 
the vacuum projector $\vert 0 \rangle \langle 0 \vert$  in hadronic operators. 
As a consequence, any matrix element involving derivatives acting on both the quark-antiquark 
fields may be reduced by integration by parts either to a matrix element that does not involve 
an operator with derivatives acting on the quark-antiquark fields or to a global derivative 
of a matrix element of the type $\langle 0 \vert (\dots) \vert H\rangle$.
The first one is a standard matrix element that does not involve the center of mass 
momentum, the last one vanishes in the quarkonium center of mass rest frame.

\subsection{Hadronic decay widths}
\label{haddecwidth-general}
Having assumed a power counting and having chosen a basis of operators, we are in the position to provide explicit
factorization formulas for S-wave and P-wave inclusive decays. 
The S-wave decay widths  at order $v^7$ are:
\bea
&& \hspace{-4mm}
\Gamma (^1S_0\rightarrow \textrm{l.h.}) = \frac{2\,{\rm Im}\,
f_1(^1S_0)}{M^2}\langle H(^1S_0) | \mathcal O_1 (^1S_0) |H(^1S_0)
\rangle 
\nn\\
&& \hspace{-4mm}
+ \frac{2\, {\rm Im}\, g_1(^1S_0)}{M^4}\langle H(^1S_0) |\mathcal P_1 (^1S_0) |H(^1S_0) \rangle 
+ \frac{2\,{\rm Im}\,f_8(^3S_1)}{M^2}\langle H(^1S_0) | \mathcal O_8 (^3S_1) |H(^1S_0) \rangle 
\nn\\
&& \hspace{-4mm}
+ \frac{2\,{\rm Im}\, f_{8}( ^1S_0) }{M^2}\langle H(^1S_0) | \mathcal O_{8}(^1S_0) |H(^1S_0) \rangle 
+ \frac{2\,{\rm Im}\, f_8(^1P_1)}{M^4}\langle H(^1S_0) | \mathcal O_8(^1P_1) |H(^1S_0) \rangle 
\nn\\
&& \hspace{-4mm}
+ \frac{2\,{\rm Im}\,s_{1\textrm{-}8}(^1S_0,^3S_1)}{M^4}\langle H(^1S_0) | \mathcal S_{1\textrm{-}8} (^1S_0,^3S_1) |H(^1S_0) \rangle 
+ \frac{2\,{\rm Im}\, f'_{8 \, \textrm{cm}}}{M^4}\langle H(^1S_0) |\mathcal O'_{8\,\textrm{cm}} |H(^1S_0) \rangle 
\nn\\ 
&& \hspace{-4mm}
+  \frac{2\,{\rm Im}\, g_{8 a \,\textrm{cm}} }{M^4}\langle H(^1S_0) | \mathcal P_{8 a \,\textrm{cm}} |H(^1S_0) \rangle 
+ \frac{2\,{\rm Im}\, f_{1 \,\textrm{cm}}}{M^4}\langle H(^1S_0) | \mathcal O_{1\,\textrm{cm}} |H(^1S_0) \rangle 
\nn\\
&& \hspace{-4mm}
+ \frac{2\,{\rm Im}\, h'_1(^1S_0)}{M^6}\langle H(^1S_0) |\mathcal Q'_1 (^1S_0) |H(^1S_0) \rangle 
+ \frac{2\,{\rm Im}\,h''_1(^1S_0)}{M^6}\langle H(^1S_0) | \mathcal Q''_1 (^1S_0) |H(^1S_0) \rangle  
\nn \\
&& \hspace{-4mm}
+ \frac{2\,{\rm Im}\, g_{8}( ^3S_1) }{M^4}\langle H(^1S_0) | \mathcal P_{8}(^3S_1) |H(^1S_0) \rangle   
+ \frac{2\,{\rm Im}\, g_{8}( ^1S_0) }{M^4}\langle H(^1S_0) | \mathcal P_{8}(^1S_0) |H(^1S_0) \rangle
\nn \\
&& \hspace{-4mm}
+ \frac{2 \textrm{Im}\, g_8(^1P_1)}{M^6} \langle H(^1S_0) | \mathcal P_8(^1P_1) | H(^1S_0) \rangle 
+ \frac{2 \textrm{Im}\, h'_8(^1S_0)}{M^6} \langle H(^1S_0) | \mathcal Q'_8(^1S_0) | H(^1S_0) \rangle 
\nn \\
&& \hspace{-4mm}
+ \frac{2 \textrm{Im}\, h_8(^1D_2)}{M^6} \langle H(^1S_0) | \mathcal Q_8(^1D_2) | H(^1S_0) \rangle 
+ \frac{2 \textrm{Im}\, h_1(^1D_2)}{M^6} \langle H(^1S_0) | \mathcal Q_1(^1D_2) | H(^1S_0) \rangle
\nn\\
&& \hspace{-4mm}
+ \frac{2 \textrm{Im}\, d_8(^1S_0,^1P_1)}{M^5} \langle H(^1S_0) | \mathcal D_{8\textrm{-}8}(^1S_0,^1P_1) | H(^1S_0) \rangle,
\label{eq:hadwids0}
\eea

\bea
&& \hspace{-4mm}
\Gamma (^3S_1\rightarrow \textrm{l.h.}) = 
\frac{2\,{\rm Im}\,f_1(^3S_1)}{M^2}\langle H(^3S_1) | \mathcal O_1 (^3S_1) |H(^3S_1) \rangle 
\nn\\
&& \hspace{-4mm}
+ \frac{2\, {\rm Im}\, g_1(^3S_1)}{M^4}\langle H(^3S_1) |\mathcal P_1 (^3S_1) |H(^3S_1) \rangle 
+ \frac{2\,{\rm Im}\,f_8(^1S_0)}{M^2}\langle H(^3S_1) | \mathcal O_8 (^1S_0) |H(^3S_1) \rangle  
\nn\\
&& \hspace{-4mm}
+ \frac{2\,{\rm Im}\,f_8(^3S_1)}{M^2}\langle H(^3S_1) | \mathcal O_8(^3S_1) |H(^3S_1) \rangle 
+ \sum_{J =0}^2\frac{2\,{\rm Im}\,f_8(^3P_J)}{M^4}\langle H(^3S_1) | \mathcal O_8(^3P_J) |H(^3S_1) \rangle 
\nn\\ 
&& \hspace{-4mm}
+ \frac{2\,{\rm Im}\, s_{1\textrm{-}8}(^3S_1,^1S_0)}{M^4}\langle H(^3S_1) | \mathcal S_{1\textrm{-}8} (^3S_1,^1S_0) |H(^3S_1)\rangle  
+ \frac{2\,{\rm Im}\, f_{8 \,\textrm{cm}}}{M^4}\langle H(^3S_1) |\mathcal O_{8\,\textrm{cm}} |H(^3S_1) \rangle 
\nn\\ 
&& \hspace{-4mm}
+  \frac{2\,{\rm Im}\, g_{8c\,\textrm{cm}} }{M^4}\langle H(^3S_1) | \mathcal P_{8c\,\textrm{cm}} |H(^3S_1) \rangle
+ \frac{2\,{\rm Im}\, f'_{1 \,\textrm{cm}}}{M^4}\langle H(^3S_1) |\mathcal O'_{1\,\textrm{cm}} |H(^3S_1) \rangle 
\nn\\
&& \hspace{-4mm}
+ \frac{2\,{\rm Im}\, h'_1(^3S_1)}{M^6}\langle H(^3S_1) |\mathcal Q'_1 (^3S_1) |H(^3S_1) \rangle 
+ \frac{2\,{\rm Im}\,h''_1(^3S_1)}{M^6}\langle H(^3S_1) | \mathcal Q''_1 (^3S_1)|H(^3S_1) \rangle  
\nn\\
&& \hspace{-4mm}
+ \frac{2\,{\rm Im}\, g_1(^3S_1,^3D_1)}{M^4}\langle H(^3S_1) | \mathcal P_1(^3S_1,^3D_1) |H(^3S_1) \rangle  
+ \frac{2\,{\rm Im}\, g_{8}( ^1S_0) }{M^4}\langle H(^3S_1) | \mathcal P_{8}(^1S_0) |H(^3S_1) \rangle 
\nn\\
&& \hspace{-4mm}
+ \frac{2\,{\rm Im}\, g_8(^3S_1)}{M^4}\langle H(^3S_1) | \mathcal P_8 (^3S_1) |H(^3S_1) \rangle 
+ \frac{2\,{\rm Im}\, t^{(1) \prime}_{1\textrm{-}8}(^3S_1,^3P) }{M^5}
\langle H(^3S_1) | \mathcal T^{(1) \prime}_{1\textrm{-}8}(^3S_1,^3P) |H(^3S_1) \rangle
\nn\\
&& \hspace{-4mm}
+ \sum_{J =0}^2 \frac{2 \textrm{Im}\, g_8(^3P_J)}{M^6} \langle H(^3S_1) | \mathcal P_8(^3P_J) | H(^3S_1) \rangle 
+ \frac{2 \textrm{Im}\, h'_8(^3S_1)}{M^6} \langle H(^3S_1) | \mathcal Q'_8(^3S_1) | H(^3S_1) \rangle 
\nn\\
&& \hspace{-4mm}
+ \sum_{J =0}^2\left[\frac{2 \textrm{Im}\, h_8(^3D_J)}{M^6} \langle H(^3S_1) | \mathcal Q_8(^3D_J) | H(^3S_1) \rangle 
+ \frac{2 \textrm{Im}\, h_1(^3D_J)}{M^6} \langle H(^3S_1) | \mathcal Q_1(^3D_J) | H(^3S_1) \rangle \right]
\nn\\
&& \hspace{-4mm}
+ \sum_{k =0,2}
\frac{2 \textrm{Im}\, d^{(k)}_8(^3S_1,^3P)}{M^5} \langle H(^3S_1) | \mathcal D^{(k)}_{8\textrm{-}8}(^3S_1,^3P) | H(^3S_1) \rangle
\nn\\
&& \hspace{-4mm}
+ \frac{2 \textrm{Im}\, g_8(^3P_2,^3F_2)}{M^6} \langle H(^3S_1) | \mathcal P_8(^3P_2,^3F_2) | H(^3S_1) \rangle.
\label{eq:hadwid3s1}
\eea
In Eqs. \eqref{eq:hadwids0} and \eqref{eq:hadwid3s1}, the first matrix element scales like $v^3$, 
the following four in the second and third line like $v^5$, the following two like $v^6$ and the others 
like $v^7$. S-wave decay widths at order $v^7$ were computed in \cite{Bodwin:2002hg}. 
For $\Gamma(^1S_0 \rightarrow \textrm{l.h.})$, the decay width in \cite{Bodwin:2002hg} does not include 
the matrix elements of the operators proportional to the total momentum of the $Q \bar Q$ pair, 
the matrix element of $ \mathcal Q_1(^1D_2)$ and any other matrix element of octet 
operators with the exception of $\mathcal O_{8}(^3S_1)$,
$\mathcal O_{8}(^1S_0)$ and $\mathcal O_{8}(^1P_1)$.
In the power counting adopted in \cite{Bodwin:2002hg}, which is described in \cite{Bodwin:1994jh}, 
all these matrix elements are suppressed by further powers of $v$
and they can be neglected at this order of the expansion.
For the same reason, the expression for $\Gamma(^3S_1 \rightarrow \textrm{l.h.})$ in \cite{Bodwin:2002hg} 
does not include all the matrix elements of operators proportional 
to the total momentum of the $Q\bar Q$ pair, 
the matrix elements of $ \mathcal Q_1(^3D_J)$ and $\mathcal P_1(^3S_1,^3D_1)$, and any other matrix element of octet 
operators with the exception of $\mathcal O_{8}(^3S_1)$, $\mathcal O_{8}(^1S_0)$ 
and $\mathcal O_{8}(^3P_J)$.

The P-wave decay widths at order $v^7$ are:
\bea
&& \hspace{-4mm}
\Gamma (^3P_J\rightarrow \textrm{l.h.}) = 
\frac{2\,{\rm Im}\,f_1(^3P_J)}{M^4}\langle H(^3P_J) | \mathcal O_1 (^3P_J) |H(^3P_J) \rangle 
\nn\\
&& \hspace{-4mm}
+ \frac{2\, {\rm Im}\, f_8(^3S_1)}{M^2}\langle H(^3P_J) |\mathcal O_8 (^3S_1) |H(^3P_J) \rangle 
+ \frac{2\,{\rm Im}\,g_1(^3P_J)}{M^6}\langle H(^3P_J) | \mathcal P_1 (^3P_J) |H(^3P_J)\rangle 
\nn\\
&&  \hspace{-4mm}
+ \frac{2\,{\rm Im}\, g_8(^3S_1)}{M^4}\langle H(^3P_J) |\mathcal P_8 (^3S_1) |H(^3P_J) \rangle 
+ \frac{2\,{\rm Im}\,g_8(^3S_1,^3D_1)}{M^4}\langle H(^3P_J) | \mathcal P_8(^3S_1,^3D_1) |H(^3P_J) \rangle   
\nn\\
&& \hspace{-4mm}
+ \frac{2\,{\rm Im}\,g_{8a \,\textrm{cm}} }{M^4}\langle H(^3P_J) | \mathcal P_{8a \,\textrm{cm}} |H(^3P_J) \rangle    
+ \frac{2\,{\rm Im}\,t_{1\textrm{-}8}(^3P_J,^3S_1)}{M^5}\langle H(^3P_J) | \mathcal T_{1\textrm{-}8} (^3P_J,^3S_1)|H(^3P_J) \rangle  
\nn\\
&&  \hspace{-4mm}
+ \frac{2\,{\rm Im}\,f_8(^1P_1)}{M^4}\langle H(^3P_J) | \mathcal O_8 (^1P_1) |H(^3P_J) \rangle 
+ \frac{2\, {\rm Im}\, f_8(^1S_0)}{M^2}\langle H(^3P_J) | \mathcal O_8 (^1S_0) |H(^3P_J) \rangle
\nn\\
&& \hspace{-4mm}
+ \frac{2\, {\rm Im}\, f_1(^1S_0)}{M^2}\langle H(^3P_J) |\mathcal O_1 (^1S_0) |H(^3P_J) \rangle 
+ \frac{2\,{\rm Im}\, f_8(^3P_J)}{M^4}\langle H(^3P_J) | \mathcal O_8 (^3P_J) |H(^3P_J)\rangle 
\nn \\
&& \hspace{-4mm}
+ \frac{2 \textrm{Im}\, h'_8(^3S_1)}{M^6} \langle H(^3P_J) | \mathcal Q'_8(^3S_1) | H(^3P_J) \rangle 
+ \frac{2 \textrm{Im}\, h'_8(^3S_1,^3D_1)}{M^6} \langle H(^3P_J) | \mathcal Q'_8(^3S_1,^3D_1) | H(^3P_J) \rangle 
\nn \\
&& \hspace{-4mm}  
+ \sum_{k=1}^{J+1} \frac{2 \textrm{Im}\, h_8(^3D_k)}{M^6} \langle H(^3P_J) | \mathcal Q_8(^3D_k) | H(^3P_J) \rangle 
+ \frac{2 \textrm{Im}\, f_1(^3S_1)}{M^2} \langle H(^3P_J) | \mathcal O_1(^3S_1) | H(^3P_J) \rangle 
\nn \\
&& \hspace{-4mm}  
+ \sum_{i=1,8}
\delta_{J2} \frac{2 \textrm{Im}\, g_i(^3P_2,^3F_2)}{M^6} \langle H(^3P_2) | \mathcal P_i(^3P_2, ^3F_2) | H(^3P_2) \rangle, 
\label{eq:hadwidth.0}
\eea
where $J = 0,1,2$.

In Eq. \eqref{eq:hadwidth.0}, the first two matrix elements scale like $v^5$, the remaining ones like $v^7$.
P-wave decay widths at order $v^7$ were computed in \cite{Huang:1997nt}, 
where the power counting of \cite{Bodwin:1994jh} was used: they appear to  
contain only the first four terms of Eq. \eqref{eq:hadwidth.0}.
It seems, however, that also by adopting the power counting of \cite{Bodwin:1994jh} at least the matrix 
elements of the operators $\mathcal P_8(^3S_1,^3D_1)$ and $\mathcal P_{8 a\, \textrm{cm}}$
should be added.

\section{Matching}
\label{matching} 
In this section, we calculate the order $\als^2$ contributions to the imaginary parts of the matching coefficients 
that appear in Eqs. \eqref{eq:hadwids0}-\eqref{eq:hadwidth.0}. The method consists
in equating (matching) the imaginary parts of scattering
amplitudes in QCD and NRQCD along the lines of \cite{Bodwin:1994jh}.

In the QCD part of the matching, the ingoing quark and the outgoing
antiquark are represented by the Dirac spinors $u(\vec p)$ and $v(\vec p)$ 
respectively, whose explicit expressions are
\be
\label{eq:13def}
u(\vec p) = \sqrt{\frac{E_p+M}{2E_p}}
\left(
\begin{array}{c}
\xi
\\
\displaystyle \frac{\vec p \cdot \vec{\sigma}}{E_p+M}\xi
\end{array}
\right),
\qquad
v(\vec p) = \sqrt{\frac{E_{p}+M}{2E_{p}}}
\left(
\begin{array}{c}
\displaystyle \frac{\vec p \cdot \vec {\sigma}}{E_{p}+M}\eta
\\
\eta
\end{array}
\right),
\ee
where $E_p = \sqrt{{\vec p}^{\,2}+M^2}$, and $\xi$ and $\eta$ are Pauli spinors.
In the NRQCD part of the matching, the ingoing  quark and the outgoing
antiquark are represented by the Pauli spinors $\xi$ and $\eta$ respectively.

We will match singlet, octet and singlet-octet transition
operators at order $\als^2$; to this purpouse we will consider both the scattering 
amplitudes $Q \bar Q \rightarrow Q \bar Q $ and $Q \bar Q \, g \rightarrow Q \bar Q$,
with no more than two gluons in the intermediate states.

In the center of mass rest frame, the energy and momentum conservation imposes the following kinematical
constraints on the scattering  $Q \bar Q \rightarrow Q \bar Q$,
\be
|\vec p| = |\vec k|, \qquad \vec p+\vec p^{\,\prime}=0, \qquad \vec k+\vec k'=0,
\ee
and on the scattering $Q \bar Q \, g \rightarrow Q \bar Q$,
\be
E_p + E_{p'}+|\vec q| = 2E_k, \qquad \vec p+\vec p^{\,\prime} +\vec q=0, \qquad \vec k+\vec k'=0,
\ee
where $\vec p$, $\vec p^{\,\prime}$ are the ingoing and $\vec k$, $\vec k'$ the outgoing quark and antiquark
momenta, while $\vec q$ is the momentum of the ingoing gluon, which is on mass shell.

The matching does not rely on any specific power counting and can be performed 
order by order in $1/M$ \cite{Manohar:1997qy}.
We will perform the matching up to order
$1/M^6$, which is the highest power in $1/M$ appearing in Eqs.
\eqref{eq:hadwids0}-\eqref{eq:hadwidth.0}. In practice, we
expand the QCD amplitude with respect to all external three-momenta. 
Note that, in the relativistic expansion, the gluon momentum $|\vec q|$ 
is proportional to (three-momenta)$^2/M$. In the matching calculation, therefore, 
the gluon three-momentum appears with an extra $1/M$ suppression 
with respect to the quark and antiquark three-momenta.
In the case of the $Q \bar Q \, g \rightarrow Q \bar Q$ scattering,
the expansion in the gluon momentum may develop infrared
singularities, i.e. terms proportional to $1/|\vec q|$. 
These terms cancel in the matching, as expected, having QCD and NRQCD the same infrared structure.
For a detailed discussion see \cite{Brambilla:2006ph}.
In the hadronic calculation, individual diagrams that contribute 
to the imaginary part of the $Q \bar Q \, g \rightarrow Q \bar Q$ scattering
amplitude containing interactions between the gluon in the initial state and gluon propagators
develop also collinear singularities, i.e. terms proportional to $1/(1\pm \cos\theta)$, 
$\theta$ being the angle between the incoming gluon momentum and the momentum flowing 
in one of the gluon propagators put on shell  to get the imaginary contribution.  
These singular terms cancel in the sum of all diagrams.
Finally, we expect that, since the matching does not rely on a power counting 
and scattering amplitudes do not have a definite angular momentum, the matching will determine
more coefficients than needed in Eqs. \eqref{eq:hadwids0}-\eqref{eq:hadwidth.0}.

\subsection{$Q \bar Q $ to light hadrons: singlet matching}
\label{QQ1}
\begin{figure}[t]
\includegraphics[width=7cm]{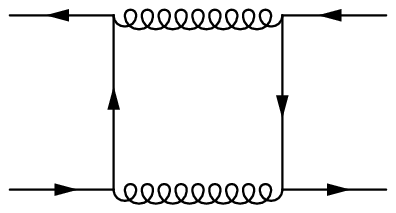}
\includegraphics[width=7cm]{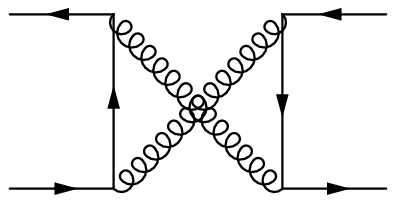}\\
\includegraphics[width=5cm]{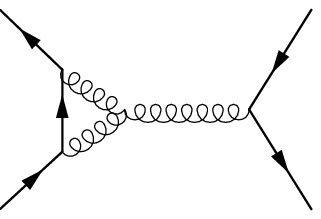}
\includegraphics[width=5cm]{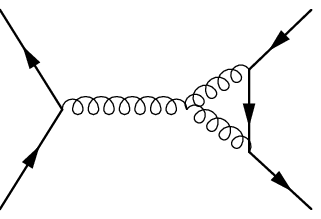}\\
\includegraphics[width=5cm]{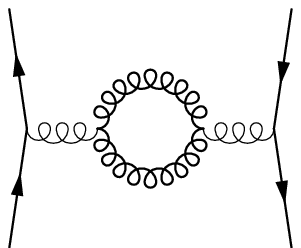}
\includegraphics[width=5cm]{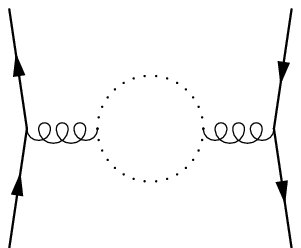}\\
\includegraphics[width=5cm]{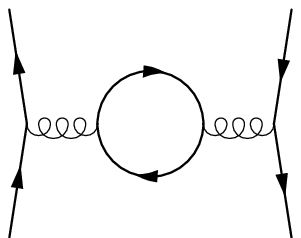}
 \caption{ \label{Fig:3} QCD Feynman diagrams describing the amplitude $Q \bar Q \to Q \bar Q$ at order $\als^2$.}
\end{figure}

\begin{figure}[ht]
\parbox{15cm}{
\centering
\includegraphics[width=3.5cm]{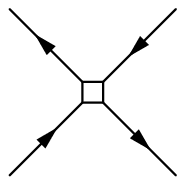}
\caption{\label{Fig:3bis} Generic NRQCD four-fermion Feynman diagram.
The empty box stands for one of the four-fermion vertices induced by the
operators listed in appendix \ref{AppA}, Eqs. \eqref{eq:B.1}-\eqref{eq:B.3oct}
and  \eqref{eq:B.4}-\eqref{eq:B.6oct}.
}
}
\end{figure}

The matching of the  $Q \bar Q \to g g \, ( q \bar q ) \to Q \bar Q$ amplitude is performed by
equating the sum of the imaginary parts of the QCD diagrams shown
in Fig. \ref{Fig:3} (taken by cutting the gluon propagators or the light quark propagators
according to $1/k^2 \to -2\pi\, i\, \delta(k^2) \theta(k^0)$) to
the sum of all the NRQCD diagrams of the type shown in Fig.
\ref{Fig:3bis}. The first two diagrams in Fig. \ref{Fig:3} contain
both a color singlet and a color octet part, coming from the
decompositions
\begin{equation}\label{eq:QQlh.1}
\begin{split}
t^a t^b \otimes t^b t^a &= \frac{C_F}{2N_c} \onec \otimes \onec +
\frac{N_c^2-2}{2N_c} t^a \otimes t^a \,,\\
t^a t^b \otimes t^a t^b &= \frac{C_F}{2N_c} \onec \otimes \onec -
\frac{1}{N_c} t^a \otimes t^a \,,
\end{split}
\end{equation}
while the other five Feynman diagrams contribute only to the octet
part.

The calculation of the box diagrams in Fig. \ref{Fig:3} gives the matching coefficients 
of the dimension 6, 8 and 10 singlet operators proportional to the relative momentum of the $Q \bar Q$ pair, 
listed in Eqs. \eqref{eq:B.1}-\eqref{eq:B.3} and  \eqref{eq:B.4}-\eqref{eq:B.6}. 
We quote the coefficients of the dimension 6 and dimension 8 operators  in appendix \ref{AppB}. They
agree with those calculated in \cite{Bodwin:1994jh}. 
We refer to \cite{Vairo:2003gh} and references therein for an updated list of imaginary parts of
matching coefficients of dimension 6 and 8 four-fermion operators; 
some of them are known at next-to-leading order. For dimension 10 operators we find  
\bea 
{\rm Im} \,h_{1}(^1D_2) &=& \frac{2}{15}\frac{\als^2 \,\pi\,
C_F}{2N_c}, \label{c5}
\\
{\rm Im} \,h'_{1}(^1S_0)+{\rm Im} \,h''_{1}(^1S_0) &=&
\frac{68}{45}\frac{\als^2 \,\pi\, C_F}{2N_c}, \label{c6}
\\
{\rm Im} \,g_{1}(^3P_0) &=& -7 \frac{\als^2 \,\pi\,
C_F}{2N_c}, \label{c7}
\\
{\rm Im} \,g_{1}(^3P_2) &=& -\frac{8}{5}\frac{\als^2 \,\pi\,
C_F}{2N_c}, \label{c8}
\\
{\rm Im} \,g_{1}(^3P_2,^3F_2) &=& -\frac{20}{21}\frac{\als^2\,
\pi \,C_F}{2N_c}. \label{c9} 
\eea 
The four-fermion operators to which the matching coefficients refer are listed in appendix \ref{AppA}.

The coefficients relevant for P-wave decay widths at order $v^7$ are \eqref{c7} and \eqref{c8}.
They were first computed in \cite{Huang:1997nt}, but our results 
disagree with the ones reported there. Note that while Eqs.  \eqref{c7} and \eqref{c8} agree in the QED limit 
with the results of \cite{Brambilla:2006ph}, the QED limit of the results in \cite{Huang:1997nt}
is in disagreement both with  \cite{Brambilla:2006ph} and \cite{Ma:2002ev}. 

Equation \eqref{c6} agrees with the one found in \cite{Bodwin:2002hg}. 
By matching the diagrams of Fig. \ref{Fig:3} we cannot resolve ${\rm Im} \,h'_{1}(^1S_0)$ 
and ${\rm Im} \,h''_{1}(^1S_0)$ separately.
These coefficients multiply operators that contribute to the $v^4$ corrections of the 
S-wave decay widths. Eq. \eqref{c5} contributes to the leading order decay width 
of the singlet state of the $D$ multiplet, which for charmonium and bottomonium 
has not yet been observed; it agrees with the result of \cite{Novikov:1977dq}.

\subsection{$Q \bar Q$ to light hadrons: octet matching}
\label{QQ8}
The calculation of the diagrams in Fig. \ref{Fig:3} provides also the
coefficients of dimension 6, 8 and 10 color octet operators.
Again, since we work in the center of mass rest frame, we cannot obtain the matching 
coefficients of the dimension 8 and 10 operators proportional to the center of mass momentum.

The coefficients of the dimension 6 and 8 operators are quoted in appendix \ref{AppB} and agree with
those obtained  in \cite{Bodwin:1994jh} and \cite{Petrelli:1997ge}.
The coefficients of the dimension 10 operators are new results of this work.
We find 
\bea
{\rm Im} \,h'_{8}(^3S_1)+{\rm Im}
\,h''_{8}(^3S_1) &=& \frac{29}{108} \als^2  \pi n_f
+ \frac{1}{108} \als^2  \pi N_c , 
\label{c13}\\
{\rm Im} \,h_{8}'(^3S_1,^3D_1)+{\rm Im}
\,h_{8}''(^3S_1,^3D_1) &=& \frac{23}{72}\als^2\pi
n_f + \frac{1}{18}\als^2 \pi N_c,
\label{c17} \\
{\rm Im} \,h_{8}(^3D_1) &=&\frac{1}{24}\als^2 \pi
n_f + \frac{1}{12} \als^2  \pi N_c, 
\label{c14} \\
{\rm Im} \,h_{8}(^3D_2) &=& \frac{1}{30}\als^2 \pi N_c , 
\label{c15} \\
{\rm Im} \,h_{8}(^3D_3) &=& \frac{1}{21} \als^2 \pi N_c , 
\label{c16} \\
{\rm Im} \,h_{8}(^1D_2) &=& \frac{2}{15}\als^2 \,\pi
\frac{N^2_c - 4 }{4N_c}, 
\label{c23} \\
{\rm Im} \,h'_{8}(^1S_0)+{\rm Im} \,h''_{8}(^1S_0) &=&
\frac{68}{45}\als^2 \,\pi \frac{N^2_c - 4 }{4N_c}, 
\label{c24} \\
{\rm Im} \,g_{8}(^1P_1) &=& -\frac{3}{20} \als^2 \,\pi N_c, 
\label{c25}\\
 {\rm Im} \,g_{8}(^3P_0) &=& -7 \als^2
\,\pi \frac{N^2_c - 4 }{4N_c}, 
\label{c26} \\
{\rm Im} \,g_{8}(^3P_2) &=& -\frac{8}{5}\als^2 \,\pi
\frac{N^2_c - 4 }{4N_c}, 
\label{c27} \\
{\rm Im} \,g_{8}(^3P_2,^3F_2) &=& -\frac{20}{21}\als^2 \,\pi
\frac{N^2_c - 4 }{4N_c} \,.
\label{c28} 
\eea
The four-fermion operators to which the matching coefficients
refer are listed in appendix \ref{AppA}.

\subsection{$Q \bar Q \, g$ to light hadrons}
\label{QQg}
\begin{figure}[ht]
\parbox{15cm}{
\centering
\includegraphics[width=6.5cm]{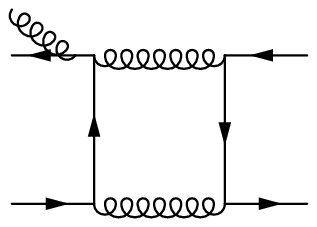}
\includegraphics[width=6.5cm]{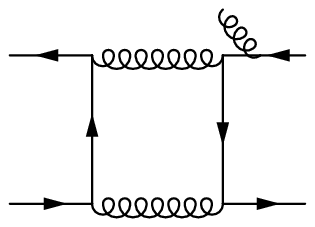}
\vspace{1.5cm}\\
\includegraphics[width=6.5cm]{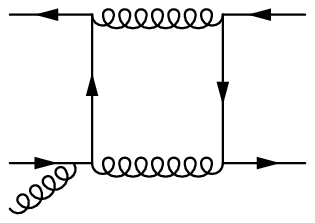}
\includegraphics[width=6.5cm]{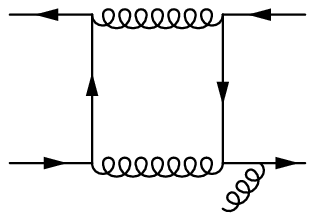}
\includegraphics[width=6.5cm]{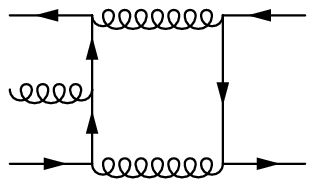}
\includegraphics[width=6.5cm]{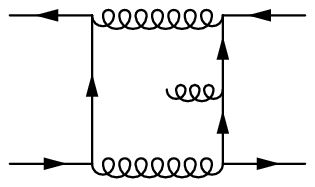}
\caption{\label{Fig:4} Box Diagrams: the gluon in the initial state interacts with a fermion leg. 
The other six diagrams, in which the two gluon propagators cross, have not been displayed.}
}
\end{figure}

\begin{figure}[ht]
\parbox{15cm}{
\centering
\includegraphics[width=6.5cm]{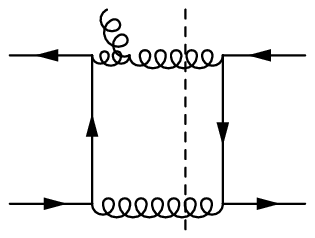}
\includegraphics[width=6.5cm]{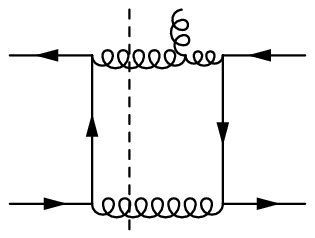}
\includegraphics[width=6.5cm]{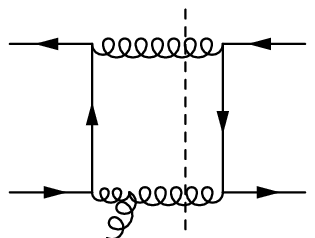}
\includegraphics[width=6.5cm]{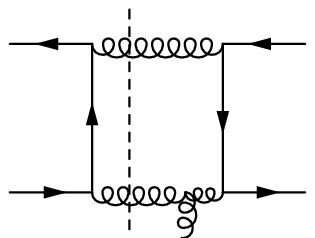}
\caption{\label{Fig:5} Box Diagrams: the gluon in the initial state interacts with a gluon propagator. 
The other four diagrams, in which the two gluon propagators cross, have not been displayed.}
}\end{figure}

\begin{figure}[ht]
\parbox{15cm}{
\centering
\includegraphics[width=5cm]{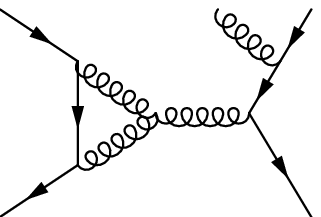} \hspace{1cm} 
\includegraphics[width=5cm]{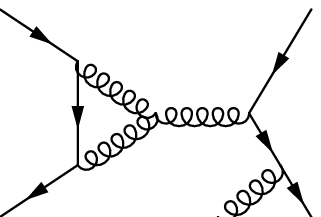} \vspace{0.3cm} \\
\includegraphics[width=5cm]{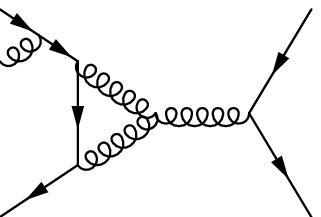} \hspace{1cm}
\includegraphics[width=5cm]{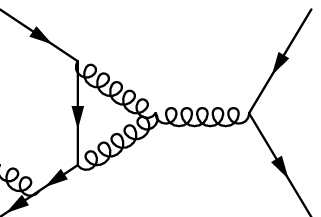} \vspace{0.3cm} \\
\includegraphics[width=5cm]{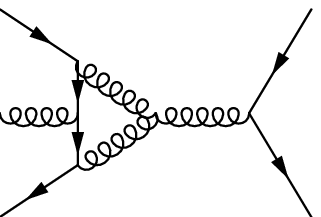} \hspace{1cm}
\includegraphics[width=5cm]{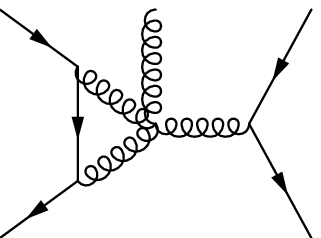} 
\caption{\label{Fig:6} Vertex corrections: the gluon in the initial state couples to a fermion leg or to a three gluon vertex.}
}\end{figure}

\begin{figure}[ht]
\parbox{15cm}{
\centering
\includegraphics[width=5cm]{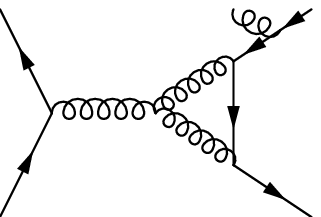} \hspace{1cm}
\includegraphics[width=5cm]{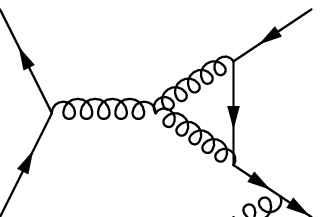} \vspace{0.3cm} \\
\includegraphics[width=5cm]{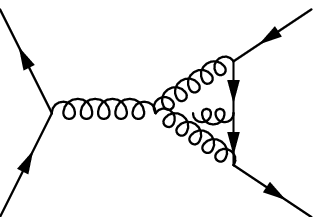} \hspace{1cm}
\includegraphics[width=5cm]{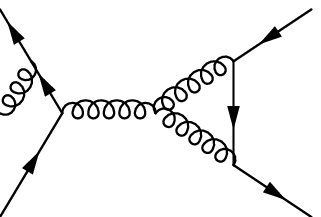} \vspace{0.3cm} \\
\includegraphics[width=5cm]{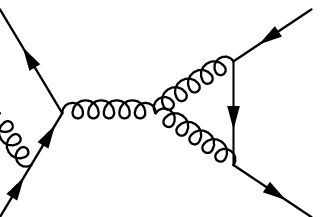} \hspace{1cm}
\includegraphics[width=5cm]{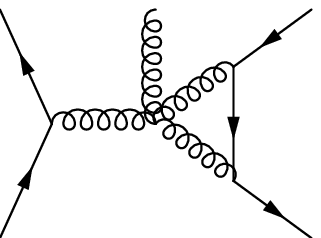} 
\caption{\label{Fig:7} Vertex corrections: the gluon in the initial state couples to a fermion leg or to a three gluon vertex.}
}\end{figure}

\begin{figure}[ht]
\parbox{15cm}{
\centering
\includegraphics[width=5cm]{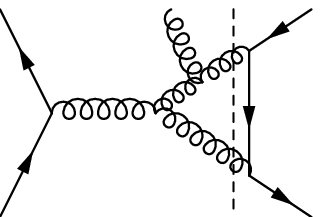} \hspace{1cm}
\includegraphics[width=5cm]{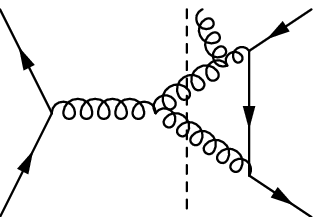} \vspace{.3cm} \\
\includegraphics[width=5cm]{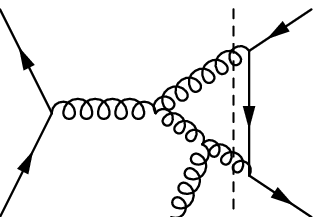} \hspace{1cm}
\includegraphics[width=5cm]{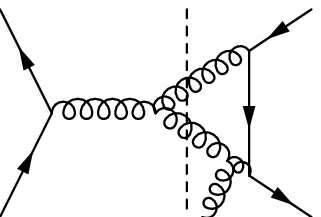} \vspace{.3cm} \\
\includegraphics[width=5cm]{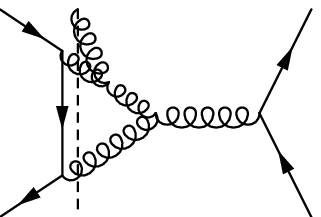} \hspace{1cm}
\includegraphics[width=5cm]{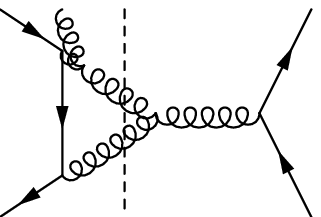} \vspace{.3cm} \\
\includegraphics[width=5cm]{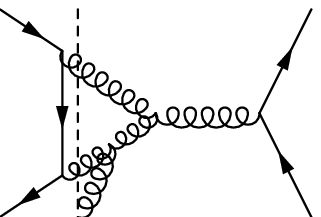} \hspace{1cm}
\includegraphics[width=5cm]{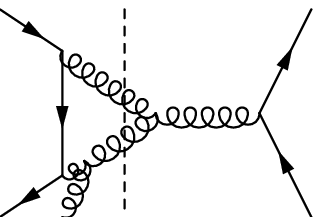}
\caption{\label{Fig:8} Vertex corrections: the gluon in the initial state interacts with a gluon propagator.}
}\end{figure}

\begin{figure}[ht]
\parbox{15cm}{
\centering
\includegraphics[width=5cm]{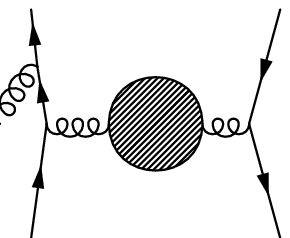} \hspace{1cm}
\includegraphics[width=5cm]{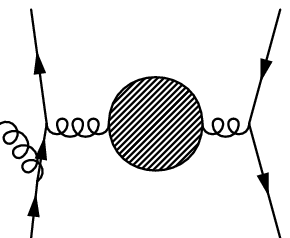} \vspace{0.5cm} \\
\includegraphics[width=5cm]{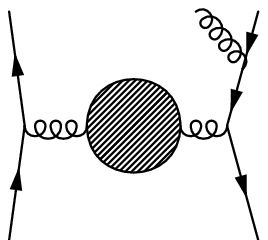} \hspace{1cm}
\includegraphics[width=5cm]{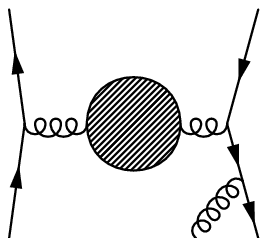} \vspace{0.5cm} \\
\includegraphics[width=3cm]{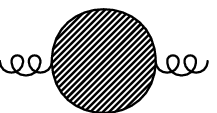} \raisebox{0.65cm}{ = }
\includegraphics[width=3cm]{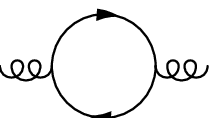} \hspace{.25cm} \includegraphics[width=3cm]{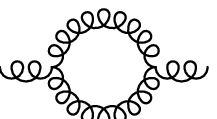}  
\hspace{.25cm} \includegraphics[width=3cm]{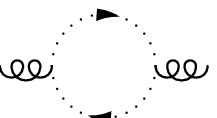}  
\caption{\label{Fig:9} Vacuum polarization: the gluon in the initial state interacts with a fermion leg.}
}\end{figure}

We show in figures \ref{Fig:4}-\ref{Fig:9} the diagrams
that  contribute to the $Q\bar Q g  \to Q \bar Q$ scattering
amplitude with terms with  color content $t^a \otimes \mathbf{1}$ or $\mathbf{1} \otimes t^a$.
The imaginary part of $Q \bar Q g \rightarrow Q \bar Q$
is computed considering all possible cuts of the gluon propagators.
Diagrams in Fig. \ref{Fig:5} and Fig. \ref{Fig:8} develop collinear singularities, 
that however cancel when all possible cuts are taken into
account. In these figures, the cuts are explicitly indicated.

\begin{figure}[ht]
\parbox{15cm}{
\centering
\includegraphics[width=2.9cm]{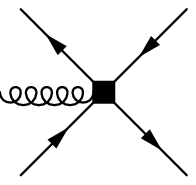}
\includegraphics[width=2.9cm]{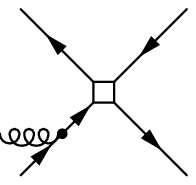}
\includegraphics[width=2.9cm]{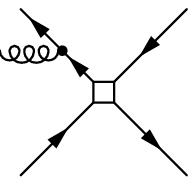}
\includegraphics[width=2.9cm]{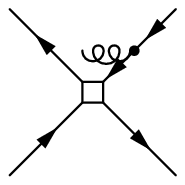}
\includegraphics[width=2.9cm]{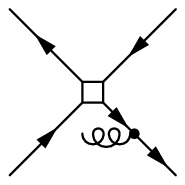}
\caption{\label{Fig:4bis} Generic NRQCD four-fermion Feynman diagrams involving an ingoing
$Q \bar Q$ pair and a gluon and an outgoing $Q \bar Q$ pair.
The black box with a gluon attached to it and the empty box
stand respectively for one of the four-fermion-one-gluon vertices
and for one of the four-fermion vertices induced by the
operators listed in appendix \ref{AppA}, Eqs. \eqref{eq:B.1}-\eqref{eq:B.6oct}.
The black dot with a gluon attached to it stands for one of the quark-gluon vertices induced by the
bilinear part of the NRQCD Lagrangian given in Eq. \eqref{NRQCD:bilinear}.}
}
\end{figure}

In the matching procedure, the QCD amplitude is equated to
the sum of all NRQCD diagrams of the type shown in Fig. \ref{Fig:4bis}.
These are all diagrams of NRQCD with an ingoing $Q \bar Q$ pair and a gluon
and an outgoing $Q \bar Q$ pair. They can involve four-fermion
operators and a gluon coupled to the quark or the antiquark line, but also
four-fermion operators that couple to gluons. Four-fermion operators 
that induce octet to singlet transitions on the $Q \bar Q$ pair may be one of the
operators listed in Eqs. \eqref{eq:B.8} and \eqref{eq:B.9}, but also one of
the four-fermion operators involving only covariant derivatives, which,
despite being usually denoted as singlet (or octet) operators, couple to the gluon field
through the term $-i t^a \, g \vec A^a$ in the covariant derivative and therefore have a singlet-octet component.

The calculation of the imaginary part of the $Q\bar Q g \rightarrow Q \bar Q$ 
scattering amplitude allows to find the matching coefficients of dimension 8 and dimension 9 
singlet-octet transition operators and dimension 8 operators proportional 
to the total momentum of the $Q \bar Q$ pair. 
It also allows to fix the individual coefficients appearing in Eqs. \eqref{c6}  and \eqref{c24}.
As discussed in section \ref{redef}, the basis of operators that we chose contains as independent
 operators $\mathcal T_{1\textrm{-}8}(^3P_J,^3S_1)$, with $J=0,1,2$. 
In appendix \ref{AppB}, we give for completeness also the matching coefficients computed with 
the other possible choice of independent operators, 
$1/2\left(\mathcal Q'_8(^3S_1) - \mathcal Q''_8(^3S_1) \right) $, 
 $1/2 \left(\mathcal Q'_8(^3S_1,^3D_1) - \mathcal Q''_8(^3S_1,^3D_1) \right) $ 
and   $\mathcal T^{(1)\prime}_{8\textrm{-}1}(^3S_1,^3P)$. 
This second set allows to establish the individual coefficients of the operators appearing 
in Eqs. \eqref{c13} and \eqref{c17}, but it is less useful for the discussion of the P-wave decay widths.  

The matching coefficients are:
\bea
{\rm Im} \,s_{1\textrm{-}8}(^1S_0,^3S_1) & =& - \frac{1}{4}\als^2 \pi + \frac{1}{12} \frac{\als^2 \pi n_f}{N_c},
\label{c29}
\\
{\rm Im} \,s_{1\textrm-8}(^3S_1,^1S_0) & =& 0,
\label{c30}
\\
{\rm Im} \,h\,'_{1}(^1S_0) &=& \frac{10}{9} \als^2 \pi \frac{C_F}{2N_c} -\frac{1}{48} \als^2 \pi,
\label{c31}
\\
{\rm Im} \,h\,''_{1}(^1S_0) &=& \frac{2}{5}\als^2 \pi \frac{C_F}{2N_c} + \frac{1}{48} \als^2 \pi ,
\label{c32}
\\
{\rm Im} \,h\,'_{8}(^1S_0) &=& \frac{10}{9} \als^2 \pi \frac{N^2_c-4}{4N_c},
\label{c33}
\\
{\rm Im} \,h\,''_{8}(^1S_0) &=& \frac{2}{5}\als^2 \pi \frac{N^2_c-4}{4N_c},
\label{c34}
\eea
\bea
{\rm Im} \,t^{(1)\prime}_{1\textrm-8}(^3S_1,^3P) & =& - \frac{1}{8} \als^2 \pi \frac{N^2_c-4}{4N_c^2},
\label{c35}
\\
{\rm Im} \,h'_{1}(^3S_1) & =& \frac{1}{12} \als^2 \pi \frac{N^2_c-4}{4N_c^2},
\label{c36}
\\
{\rm Im} \,h''_{1}(^3S_1) & =& -\frac{1}{12} \als^2 \pi \frac{N^2_c-4}{4N_c^2},
\label{c37}
\\
{\rm Im} \,h'_{1}(^3S_1,^3D_1) & =& \frac{1}{4} \als^2 \pi \frac{N^2_c-4}{4N_c^2} ,
\label{c38}
\\
{\rm Im} \,h''_{1}(^3S_1,^3D_1) & =& -\frac{1}{4} \als^2 \pi \frac{N^2_c-4}{4N_c^2} ,
\label{c39}
\eea
\bea
{\rm Im} \,t_{1\textrm-8}(^3P_0,^3S_1) &=& -\frac{3}{2}\als^2 \pi \frac{C_F}{2 N_c}  
+ \left(\frac{61}{240} + \frac{7}{192}\frac{n_f}{N_c} \right)\als^2 \pi  ,
\label{c45}
\\
{\rm Im} \,t_{1\textrm-8}(^3P_1,^3S_1) &=& \left(\frac{1}{72} + \frac{107}{576} \frac{n_f}{N_c}  \right) \als^2 \pi,
\label{c46}
\\
{\rm Im} \,t_{1\textrm-8}(^3P_2,^3S_1) &=& \left(\frac{1}{10} + \frac{25}{576} \frac{n_f}{N_c}  \right) \als^2 \pi.
\label{c47}
\eea
The four-fermion operators to which the matching coefficients refer are listed in appendix \ref{AppA}.
The total momentum of the ingoing $Q \bar Q$ being different from 0, the matching calculation for 
$Q \bar Q g \rightarrow Q \bar Q$ also provides the coefficients for 
the operators defined in \eqref{eq:B.11} and \eqref{eq:B.12}:
\bea
{\rm Im} \,f_{1\,\textrm{cm}} &=& \frac{1}{4}\als^2 \pi \frac{C_F}{2N_c},
\label{c48}
\\
{\rm Im} \,f'_{1\,\textrm{cm}} &=& 0,
\label{c49}
\\
{\rm Im} \,f_{8\,\textrm{cm}} &=& \frac{1}{4} \als^2 \pi \frac{N^2_c-4}{4N_c},
\label{c50}
\\
{\rm Im} \,f'_{8\,\textrm{cm}} &=& \frac{1}{24}\als^2 \pi n_f,
\label{c51}
\eea

\bea
{\rm Im} \,g_{1a\,\textrm{cm}} &=& 0,
\label{c52}
\\
{\rm Im} \,g_{1b\,\textrm{cm}} &=& 0, 
\label{c53}
\\
{\rm Im} \,g_{1c\,\textrm{cm}} &=& -\frac{1}{4}\als^2 \pi \frac{C_F}{2 N_c},
\label{c54}
\\
{\rm Im} \,g_{8a\,\textrm{cm}} &=& -\frac{1}{24}\als^2 \pi n_f,
\label{c55}
\\
{\rm Im} \,g_{8b\,\textrm{cm}} &=& 0, 
\label{c56}
\\
{\rm Im} \,g_{8c\,\textrm{cm}} &=& -\frac{1}{4} \als^2 \pi \frac{N^2_c-4}{4N_c}. 
\label{c57} 
\eea 
We have checked the matching coefficients \eqref{c48}-\eqref{c57} by repeating the calculation of the diagrams
in Fig. \ref{Fig:3} up to order $1/M^4$ in the general frame
\begin{equation*}
\begin{split}
\vec p = \frac{1}{2} \vec q + \vec p_r,  \qquad & \qquad \vec k = \frac{1}{2} \vec q + \vec k_r, \\
\vec p^{\,\prime} = \frac{1}{2} \vec q - \vec p_r,  \qquad & \qquad \vec k^{\, \prime} 
= \frac{1}{2} \vec q - \vec k_r. 
\end{split}
\end{equation*}
Equations \eqref{c29}-\eqref{c57} are original results of this work.

\section{Poincar\'{e} invariance constraints}
\label{poincare}
We can use Poincar\'e symmetry to obtain independent checks on some of the matching 
coefficients derived in the previous sections. Here we outline the procedure, following 
the method of Ref.~\cite{Brambilla:2003nt}.

NRQCD is constructed by expanding (and matching) QCD in the non-relativistic limit.
As a consequence, while translations and rotations are still explicit symmetries of the NRQCD action, 
the explicit invariance of the QCD action under boost is lost in the non-relativistic regime. 
However, the boost invariance of QCD manifests itself in a nonlinear realization, 
constraining the form of the NRQCD Hamiltonian. 

The constraints posed by Poincar\'{e} invariance on the bilinear sector of the NRQCD Lagrangian have been studied
extensively in \cite{Manohar:1997qy} and \cite{Brambilla:2003nt}. 
The computation of the matching coefficients \eqref{c48}-\eqref{c57} completes our knowledge of the imaginary part
of the NRQCD Lagrangian at order $1/M^4$, including four-fermion operators 
proportional to the total momentum of the $Q \bar Q$ pair, which, due to their suppression 
in $v$ have not been considered before. Therefore, we can now study  the constraints  induced 
by Poincar\'{e} invariance in  the four-fermion sector of the NRQCD Lagrangian.
We adopt the method described in \cite{Brambilla:2003nt} by constructing 
the generators of time translation $H$, space translations $\vec P$,
rotations $\vec J$ and boosts $\vec K$ inside the effective theory and 
by imposing that the  commutation relations of the Poincar\'{e} algebra
are respected. Since rotation and translation invariance are manifestly maintained in NRQCD, 
the commutation relations involving only $H$, $\vec P$ and $\vec J$ 
are trivially satisfied while the commutation relations 
involving the boost generators $\vec K$ impose restrictions among the 
matching coefficients:
\begin{eqnarray}
\left[P^i,K^j \right]  = & - i \delta^{ij} H, 
\label{eq:Poin.a6}\\ 
\left[H, K^i\right]    = & -i P^i,
\label{eq:Poin.a7} \\ 
\left[J^i,K^j \right]  = & i \varepsilon^{ijk} K^k, 
\label{eq:Poin.a8}\\ 
\left[K^i,K^j \right]  = & -i \varepsilon^{ijk} J^k .
\label{eq:Poin.a9}
\end{eqnarray}

The construction of the generators proceeds in the following way: $\vec P$ and $\vec J$ 
can be obtained from the symmetric energy-momentum tensor \cite{Brambilla:2003nt,Vairo:2003gx}:
\bea
\hspace{-4mm}
\vec P  &=& \int d^3 x \, \psi^{\dag} \left(-i \vec{D} \right) \psi + \chi^{\dag}\left( - i \vec D \right)\chi 
+ \frac{1}{2} \left[\vec{\Pi}^a \times, \vec{B}^a\right] 
\\
\hspace{-4mm}
\vec J  &=& \int d^3 x  \, \psi^{\dag} \left( \vec x \times 
\left(- i \vec D\right) + \frac{\vec{\sigma}}{2} \right) \psi 
+ \chi^{\dag} \left( \vec x \times 
\left(- i \vec D \right) + \frac{\vec{\sigma}}{2} \right) \chi 
+ \frac{1}{2} \vec x \times [\vec{\Pi}^a \times, \vec B^a],
\eea
where $(\psi, i\psi^{\dag})$, $(\chi,i\chi^{\dag})$ and, in the $A^0 =0$ gauge,
$(A_i, \Pi^i_a = \partial \mathcal L_{\rm NRQCD}/\partial (\partial_0 A_i^a))$, 
are the pairs of canonical variables.
The NRQCD Hamiltonian density $h_{\rm NRQCD}$ can be obtained from a Legendre transformation of the Lagrangian density:
\bea
H_{\rm NRQCD} &=& \int d^3 x \, h_{\rm NRQCD} 
\nn\\ 
&=&  \int d^3x \, 
\psi^\dagger \left(M - c_1 \frac{{\vec D}^2}{2M}
- c_F \frac{{\vec \sigma} \cdot g {\vec B}}{2 M} \right) \psi 
+ \chi^\dagger \left(- M + c_1 \frac{{\vec D}^2}{2M}
+ c_F \frac{{\vec \sigma} \cdot g {\vec B}}{2 M} \right) \chi  
\nn\\
&& + \frac{1}{2} \left(\vec \Pi^a \cdot \vec \Pi^a + \vec B^a \cdot \vec B^a\right) 
- \sum_{i =1,8} \frac{1}{M^2} \left(f_i(^3S_1)\, \mathcal O_i(^3S_1) + f_i(^1S_0)\, \mathcal O_i(^1S_0) \right) 
\nn\\ 
&& - \sum_{i= 1,8} \frac{1}{M^4} \left( g_{i}(^3S_1) \mathcal P_{i}(^3S_1) + g_i(^1S_0)\, \mathcal P_i(^1S_0)
+  \ldots \right) + \ldots \; .  
\label{eq:Poin.d}
\eea
The coefficient $c_1$ is equal to 1 at all orders in $\als$, see \cite{Manohar:1997qy} and \cite{Brambilla:2003nt}. 

A way to construct $\vec K$ is to write down the most general expression
consistent with the NRQCD symmetries and to match it to the QCD boost generator, 
$\displaystyle \vec K = -t \vec P
+ \int d^3x \frac{1}{2} \left\{ \vec x, h_{\rm QCD} \right \}$. This procedure is analogous 
to the one followed in the construction of the NRQCD Lagrangian: new matching coefficients, 
typical of $\vec K$, appear. The form of $\vec K$ in NRQCD is
\begin{equation}
\label{eq:Poin.e}
\vec K = - t \vec P + \int d^3 x \frac{1}{2} \left\{ \vec x, h_{\rm NRQCD} \right\} 
- \sum_{l = 1}^\infty \int d^3 x  \frac{k_{l}}{M^l} \vec K^{(l)}.
\end{equation}
This form is chosen in analogy to the QCD boost generator and satisfies \eqref{eq:Poin.a6}. 
$\vec K^{(l)}$ contains all the possible operators with mass dimension $M^l$
that are vectors under rotation, are odd under parity and are invariant under $C$ and $T$ transformations.

We now compute the imaginary, four-fermion part of the commutator \eqref{eq:Poin.a7} at order $1/M^3$. 
To this aim, we need the bilinear NRQCD Hamiltonian at order $1/M$, the four-fermion part 
of the NRQCD Hamiltonian at order $1/M^4$, the operator $\vec K^{(1)}$ and  four-fermion operators 
in the boost generator, which first appear in $\vec K^{(4)}$.     
The form of $\vec K^{(1)}$ 
\begin{equation}
\label{eq:Poin.f}
\vec K^{(1)} =   \frac{1}{2} \psi^{\dag} \frac{\vec{\sigma}}{2} \times (- i \vec D) \psi 
          -  \frac{1}{2} \chi^{\dag} \frac{\vec{\sigma}}{2} \times (- i \vec D) \chi,
\end{equation}
and its coefficient $k_1$ were obtained in \cite{Brambilla:2003nt}, 
where it was shown that $k_1 = 1$ to all orders in $\als$. 
In $\vec K^{(4)}$, four-fermion operators like
\begin{equation*}
\vec K^{(4)} =  \frac{1}{2} \psi^{\dag} \frac{\vec{\sigma}}{2} 
\times (- i \overleftrightarrow D) \chi \chi^{\dag} \psi
\end{equation*}
appear. We do not give 
the detailed form of $\vec K^{(4)}$ since an explicit calculation shows that 
$\displaystyle 1/M^4 \int d^3x [\vec K^{(4)}(x), H  ] = \mathcal O(1/M^5)$.

Using the canonical commutation relations we find for singlet operators at order $1/M^3$:
\begin{equation}
\label{eq:Poin.g}
\begin{split}
\left[H, K^j \right] & =   \frac{1}{M^3} \int d^3x \left[ 
\left( \partial^j(\psi^{\dag} \chi) \,  \chi^{\dagger} \psi 
- \psi^{\dagger}\chi \, \partial^j (\chi^{\dag} \psi)\right)\left(
\frac{1}{2} \textrm{Im}\, f_1(^1S_0) + 2\, \textrm{Im}\, g_{1 c\, \textrm{cm}}  \right)  \right.  
\\ 
& \left. 
+ \left( \partial^j(\psi^{\dag} \sigma^i \chi) \,  \chi^{\dagger} \sigma^i \psi 
- \psi^{\dagger} \sigma^i \chi \, \partial^j (\chi^{\dag} \sigma^i \psi)\right)\left(
\frac{1}{2} \textrm{Im}\, f_1(^3S_1) + 2\, \textrm{Im}\, g_{1 a \,\textrm{cm}}  \right) \right. 
\\ 
& \left.
+ \left( \partial^i(\psi^{\dag} \sigma^i \chi)   \chi^{\dagger} \sigma^j \psi 
- \psi^{\dagger}\sigma^j \chi \, \partial^i (\chi^{\dag} \sigma^i \psi)\right)\left(
 2\, \textrm{Im}\, g_{1 b \,\textrm{cm}}  \right) \right. 
\\ 
& \left.
- i \varepsilon^{j l m} \left( \psi^{\dag} \sigma^l \overleftrightarrow \partial^m \chi \, \chi^{\dag} \psi 
			   - \psi^{\dag} \chi\, \chi^{\dag} \sigma^l \overleftrightarrow \partial^m \psi \right)
\left( \frac{1}{4} \textrm{Im}\,f_1(^1S_0) -  \textrm{Im}\, f_{1\,\textrm{cm}} \right) \right. 
\\
& \left.
- i \varepsilon^{j l m} \left( \psi^{\dag} \sigma^l  \chi \,\chi^{\dag} \overleftrightarrow \partial^m \psi 
			   - \psi^{\dag} \overleftrightarrow \partial^m \chi\, \chi^{\dag} \sigma^l  \psi \right)
\left( \frac{1}{4}\textrm{Im}\, f_1(^3S_1) - \textrm{Im}\, f'_{1\,\textrm{cm}} \right) \right] 
\\ 
& = 0,
\end{split}
\end{equation}
and for octet operators:
\begin{equation}
\label{eq:Poin.h}
\begin{split}
\left[H, K^j \right] & =   \frac{1}{M^3}\int d^3x  \left[ 
\left( \partial^j(\psi^{\dag} t^a \chi)\,  \chi^{\dagger} t^a \psi 
- \psi^{\dagger}t^a \chi \,\partial^j (\chi^{\dag} t^a \psi)\right)\left(
\frac{1}{2} \textrm{Im}\, f_8(^1S_0) + 2\,\textrm{Im}\, g_{8 c \,\textrm{cm}}  \right) \right. 
\\ 
& \left.
+ \left( \partial^j\psi^{\dag} t^a \sigma^i \chi) \, \chi^{\dagger} t^a \sigma^i \psi 
- \psi^{\dagger}t^a \sigma^i \chi \, \partial^j
(\chi^{\dag} t^a \sigma^i \psi)\right)\left(
\frac{1}{2}\textrm{Im}\, f_8(^3S_1) + 2\,\textrm{Im}\, g_{8 a \,\textrm{cm}}  \right) \right. 
\\ 
& \left.
+ \left( \partial^i(\psi^{\dag} \sigma^i t^a \chi) \,   \chi^{\dagger} \sigma^j\psi
 - \psi^{\dagger}\sigma^j\chi \, \partial^i (\chi^{\dag} t^a\sigma^i \psi)\right)
\left( 2\,\textrm{Im}\, g_{8 b \,\textrm{cm}}  \right) \right. 
\\ 
& \left.
- i \varepsilon^{j l m} \left( \psi^{\dag} t^a \sigma^l \overleftrightarrow \partial^m \chi \, \chi^{\dag} t^a \psi 
			   - \psi^{\dag} t^a \chi \, \chi^{\dag} t^a \sigma^l \overleftrightarrow \partial^m \psi \right)
\left( \frac{1}{4}\textrm{Im}\, f_8(^1S_0) - \textrm{Im}\, f_{8\,\textrm{cm}} \right) \right. 
\\
& \left.
- i \varepsilon^{j l m} \left( \psi^{\dag} t^a \sigma^l  \chi \, \chi^{\dag} t^a \overleftrightarrow \partial^m \psi 
			   - \psi^{\dag} t^a \overleftrightarrow \partial^m \chi \, \chi^{\dag} t^a \sigma^l  \psi \right)
\!\left( \frac{1}{4}\textrm{Im}\, f_8(^3S_1) -\textrm{Im}\, f'_{8\, \textrm{cm}} \right) 
\right]  
\\
& = 0.
\end{split}
\end{equation}
Equations \eqref{eq:Poin.g} and \eqref{eq:Poin.h} imply that
\begin{equation}
\label{eq:Poin.1}
\begin{split}
\textrm{Im}\, g_{1c\, \textrm{cm}}  & = -\frac{1}{4}  \textrm{Im}\,f_1(^1S_0) ,
\\
  \textrm{Im}\, g_{8c\, \textrm{cm}}  &= -\frac{1}{4}  \textrm{Im}\,f_8(^1S_0),
\\
  \textrm{Im}\,g_{1a \, \textrm{cm}} & = -\frac{1}{4}  \textrm{Im}\,f_1(^3S_1),
\\
  \textrm{Im}\,g_{8a\, \textrm{cm}} & = -\frac{1}{4}  \textrm{Im}\,f_8(^3S_1),
\\
  \textrm{Im}\,g_{1 b \, \textrm{cm}} = \textrm{Im}\,g_{8 b \, \textrm{cm}} & = 0,
\end{split}
\end{equation}

\begin{equation}\label{eq:Poin.2}
\begin{split}
 \textrm{Im}\,f_{1 \,\textrm{cm}}  &= \frac{1}{4} \textrm{Im}\,f_1{(^1S_0)},
\\
 \textrm{Im}\,f^{\prime}_{1 \,\textrm{cm}} &= \frac{1}{4} \textrm{Im}\,f_1{(^3S_1)}, 
\\
 \textrm{Im}\,f_{8 \,\textrm{cm}}  &= \frac{1}{4} \textrm{Im}\,f_8{(^1S_0)}, 
\\
 \textrm{Im}\,f^{\prime}_{8 \,\textrm{cm}} &= \frac{1}{4} \textrm{Im}\,f_8{(^3S_1)}.
\end{split}
\end{equation}
Relations of the same form as Eqs. \eqref{eq:Poin.1} and \eqref{eq:Poin.2} hold also 
for the matching coefficients of the electromagnetic operators.  
Equations \eqref{eq:Poin.1} and \eqref{eq:Poin.2} imply that the knowledge of the
imaginary part of matching coefficients of the dimension 6 
operators completely determines the imaginary part of the coefficients of the
operators defined in Eqs. \eqref{eq:B.11} and \eqref{eq:B.12}, proportional to the 
total momentum of the $Q\bar Q$ pair. 
The coefficients \eqref{c48}-\eqref{c57}, obtained in the previous section, 
satisfy Eqs. \eqref{eq:Poin.1} and \eqref{eq:Poin.2}.

\section{Summary and Outlook}
\label{conclusions}
In the paper, we have calculated the hadronic inclusive quarkonium decay widths in NRQCD 
at order $v^7$ in the relativistic expansion and at order $\als^2$.
The electromagnetic S- and P-wave decay widths have been previously calculated at 
order $v^7$ in \cite{Bodwin:2002hg,Ma:2002ev,Brambilla:2006ph}.
If we count $\als(M) \sim v^2$, terms of order  $\als(M)^2 v^7$ are 
part of the next-to-next-to leading order (NNLO) corrections to the pseudoscalar S-wave decays and
part of the NLO corrections to the vector S-wave and the P-wave hadronic decays.

The results for the S-wave hadronic decay widths are given in Eqs.~\eqref{eq:hadwids0} and \eqref{eq:hadwid3s1}
with the coefficients at order $\als^2$  listed in appendix \ref{AppB}.
Let us first consider S-wave vector decays.
In the power counting of \cite{Bodwin:1994jh}, those coefficients together with previous results, 
including contributions to the decay width coming from three-gluon decays and loop corrections 
\cite{Bodwin:1994jh,Petrelli:1997ge,Mackenzie:1981sf,Huang:1996fa,Gremm:1997dq,Campbell:2007ws},
provide us with the full NLO expression of the hadronic inclusive decay widths, i.e. with the full 
expression up to order $\als^4 v^3$, $\als^3 v^5$ and $\als^2 v^7$.  
In the more conservative power counting adopted here, the octet terms $\displaystyle 
\sum_{k =0,2} \frac{2 \textrm{Im}\, d^{(k)}_8(^3S_1,^3P)}{M^5}$  
$\times \langle H(^3S_1) | \mathcal D^{(k)}_{8\textrm{-}8}(^3S_1,^3P) | H(^3S_1) \rangle$ need to be included. 
The matching coefficients $d^{(k)}_8(^3S_1,$ $^3P)$ are however unknown.
In the case of S-wave pseudoscalar decays, the largest uncertainties in the decay width come 
from the  NNLO correction in $\als$ to the matching coefficient ${\rm Im}\, f_1(^1S_0)$, 
from the NLO correction in $\als$ to the coefficient ${\rm Im} \,g_1(^1S_0)$, and from the $\als^2$ expression of 
${\rm Im} \,d_8(^1S_0,^1P_1)$, which are all unknown. 
If we count $\als(M) \sim v^2$, these are the only missing ingredient to complete 
the NNLO corrections to the pseudoscalar decay widths.
Note that to complete the NNLO corrections to the pseudoscalar width, 
the NNLO expression of ${\rm Im} \,f_1(^1S_0)$ and the NLO expression of ${\rm Im}\, g_1(^1S_0)$
would be necessary also in the power counting of \cite{Bodwin:1994jh}.
We recall that matching amplitudes with loops, like those required for calculating 
${\rm Im} \,f_1(^1S_0)$ and ${\rm Im}\, g_1(^1S_0)$ at NNLO and NLO respectively, 
and with two external gluons, like those required for calculating the ${\rm Im} \, d_8$ coefficients, 
have been beyond the scope of this work.

The result for the P-wave hadronic decay width, calculated up to order $\als(M)^2 v^7$, 
is given in Eq. \eqref{eq:hadwidth.0} with the coefficients at order $\als^2$ given in appendix \ref{AppB}.
In the case of P-wave vector decays, the present calculation together with previous results,
including contributions to the decay width coming from three-gluon decays and loop corrections 
\cite{Petrelli:1997ge, Hagiwara:1980nv}, provides us with the full expression 
of the hadronic inclusive decay widths up to order $ \als^3 v^5$ and $\als^2 v^7$. Explicitly we have:
\bea
&& \Gamma (^3P_0\rightarrow \textrm{l.h.}) = 
\frac{4}{3}\frac{\als^2(2M) \pi}{ M^4}\left[ 1 + \frac{\als}{\pi} 
\left( \frac{343}{27} + \frac{5}{16} \pi - \frac{58}{81} n_f   \right)\right]   
\langle H(^3P_0) | \mathcal O_1 (^3P_0) |H(^3P_0) \rangle 
\nn\\
&&
+   \frac{\als^2 (2M) \pi n_f}{3 M^2}\left[ 1  
+ \frac{\als}{\pi}\left( \frac{107}{6} + 2 \log 2 - \frac{3}{4} \pi^2 - \frac{5}{9} n_f    
+ \left(- \frac{73}{4} + \frac{67}{36} \pi^2 \right) \frac{5}{n_f} \right) \right]
\nn\\
&&
\hspace{7cm} \times \langle H(^3P_0) |\mathcal O_8 (^3S_1) |H(^3P_0) \rangle  
\nn\\
&&
- \frac{28}{9}\frac{\als^2 \pi}{M^6}\langle H(^3P_0) | \mathcal P_1 (^3P_0) |H(^3P_0)\rangle 
- \frac{4}{9} \frac{\als^2 \pi n_f}{ M^4}\langle H(^3P_0) |\mathcal P_8 (^3S_1) |H(^3P_0) \rangle 
\nn\\
&&
- \frac{\als^2 \pi n_f}{3 M^4}\langle H(^3P_0) | \mathcal P_8(^3S_1,^3D_1) |H(^3P_0) \rangle  
- \frac{ \als^2 \pi n_f }{12 M^4}\langle H(^3P_0) | \mathcal P_{8 a\,\textrm{cm}} |H(^3P_0) \rangle    
\nn\\
&&
+ \frac{\als^2 \pi}{M^5}\left(-\frac{19}{120} + \frac{7}{288}n_f\right)
\langle H(^3P_0) | \mathcal T_{1\textrm{-}8} (^3P_0,^3S_1)|H(^3P_0) \rangle 
\nn\\
&&
+ \frac{1}{2}\frac{ \als^2 \pi}{ M^4}\langle H(^3P_0) | \mathcal O_8 (^1P_1) |H(^3P_0)\rangle 
+ \frac{5}{6}\frac{\als^2 \pi}{ M^2}\langle H(^3P_0) |\mathcal O_8 (^1S_0) |H(^3P_0) \rangle
\nn\\
&&
+ \frac{4}{9} \frac{\als^2 \pi}{M^2}\langle H(^3P_0) |\mathcal O_1 (^1S_0) |H(^3P_0) \rangle 
+ \frac{5}{2}\frac{\als^2 \pi}{M^4} \langle H(^3P_0) | \mathcal O_8 (^3P_0) |H(^3P_0) \rangle 
\nn\\
&& 
+\frac{ \als^2 \pi}{ M^6}\left( - \frac{2}{3} + \frac{23}{54} n_f \right)
\langle H(^3P_0) | \mathcal Q'_8(^3S_1) | H(^3P_0) \rangle 
\nn\\
&& 
+\frac{ \als^2 \pi}{ M^6}\left(- \frac{1}{30}  + \frac{5}{9} n_f \right)
\langle H(^3P_0) | \mathcal Q'_8(^3S_1,^3D_1) | H(^3P_0) \rangle 
\nn \\
&& 
+\frac{ \als^2 \pi}{ M^6}\left(\frac{1}{2} + \frac{1}{12} n_f \right)
\langle H(^3P_0) | \mathcal Q_8(^3D_1) | H(^3P_0) \rangle \,,
\label{eq:3p0width}
\\
\nn &&
\\
\nn &&
\\
&& \Gamma (^3P_1\rightarrow \textrm{l.h.}) =  \frac{2 \als^3(2M)}{M^4} 
\left[ \left( \frac{587}{81}- \frac{317}{432} \pi^2 \right) - \frac{32}{243} n_f  \right] 
\langle H(^3P_1) | \mathcal O(^3P_1) | H(^3P_1) \rangle  
\nn\\
&&
+  \frac{\als^2 (2M) \pi n_f}{3 M^2}\left[ 1  
+ \frac{\als}{\pi}\left( \frac{107}{6} + 2 \log 2 - \frac{3}{4} \pi^2 - \frac{5}{9} n_f     
+ \left(- \frac{73}{4} + \frac{67}{36} \pi^2 \right) \frac{5}{n_f} \right) \right]
\nn\\
&&
\hspace{7cm} \times \langle H(^3P_1) |\mathcal O_8 (^3S_1) |H(^3P_1) \rangle 
\nn\\
&&
- \frac{4}{9} \frac{\als^2 \pi n_f}{ M^4}\langle H(^3P_1) |\mathcal P_8 (^3S_1) |H(^3P_1) \rangle 
- \frac{\als^2 \pi n_f}{3 M^4}\langle H(^3P_1) | \mathcal P_8(^3S_1,^3D_1) |H(^3P_1) \rangle  
\nn\\
&&
- \frac{ \als^2 \pi n_f }{12 M^4}\langle H(^3P_1) | \mathcal P_{8 a\,\textrm{cm}} |H(^3P_1) \rangle    
+ \frac{\als^2 \pi}{M^5}\left(\frac{1}{36} + \frac{107}{864}n_f\right)
  \langle H(^3P_1) | \mathcal T_{1\textrm{-}8} (^3P_1,^3S_1)|H(^3P_1) \rangle
\nn\\
&&
+ \frac{1}{2}\frac{\als^2 \pi}{ M^4}\langle H(^3P_1) | \mathcal O_8 (^1P_1) |H(^3P_1)\rangle   
+ \frac{5}{6}\frac{\als^2 \pi}{ M^2}\langle H(^3P_1) |\mathcal O_8 (^1S_0) |H(^3P_1) \rangle
\nn\\
&&
+ \frac{4}{9} \frac{\als^2 \pi}{M^2}\langle H(^3P_1) |\mathcal O_1 (^1S_0) |H(^3P_1) \rangle  
+ \frac{ \als^2 \pi}{ M^6}\left( - \frac{2}{3} +  \frac{23}{54} n_f \right)
  \langle H(^3P_1) | \mathcal Q'_8(^3S_1) | H(^3P_1) \rangle 
\nn \\
&& 
+\frac{ \als^2 \pi}{ M^6}\left(- \frac{1}{30}  + \frac{5}{9} n_f \right)
 \langle H(^3P_1) | \mathcal Q'_8(^3S_1,^3D_1) | H(^3P_1) \rangle 
\nn \\
&& 
+\frac{ \als^2 \pi}{ M^6}\left(\frac{1}{2} + \frac{1}{12} n_f \right)
  \langle H(^3P_1) | \mathcal Q_8(^3D_1) | H(^3P_1) \rangle 
\nn \\
&& 
+ \frac{1}{5} \frac{\als^2 \pi}{M^6}\langle H(^3P_1) | \mathcal Q_8(^3D_2) | H(^3P_1) \rangle 
\,,
\label{eq:3p1width}
\\
\nn &&
\\
\nn &&
\\
&& \Gamma (^3P_2\rightarrow \textrm{l.h.}) = \frac{16}{45}\frac{
\als^2 (2M) \pi}{ M^4} \left[ 1 + \frac{\als}{\pi} \left(
\frac{1801}{72} - \frac{337}{128} \pi^2 + 5 \log 2 - \frac{29}{27}n_f \right)\right]  
\nn\\
&&
\hspace{7cm} \times \langle H(^3P_2) | \mathcal O_1 (^3P_2) |H(^3P_2)\rangle 
\nn\\
&&
+ \frac{\als^2 \pi n_f}{3 M^2}\left[ 1  + \frac{\als(2M)}{\pi}\left( \frac{107}{6} 
+ 2 \log 2 - \frac{3}{4} \pi^2 - \frac{5}{9} n_f  
+ \left(- \frac{73}{4} + \frac{67}{36} \pi^2 \right) \frac{5}{n_f} \right) \right] 
\nn\\
&&
\hspace{7cm} \times \langle H(^3P_2) |\mathcal O_8 (^3S_1) |H(^3P_2) \rangle 
\nn\\
&&
- \frac{32}{45}\frac{\als^2 \pi}{M^6}\langle H(^3P_2) | \mathcal P_1 (^3P_2) |H(^3P_2)\rangle 
- \frac{4}{9} \frac{\als^2 \pi n_f}{ M^4}\langle H(^3P_2) |\mathcal P_8 (^3S_1) |H(^3P_2) \rangle 
\nn\\
&&
- \frac{\als^2 \pi n_f}{3 M^4}\langle H(^3P_2) | \mathcal P_8(^3S_1,^3D_1) |H(^3P_2) \rangle  
- \frac{ \als^2 \pi n_f }{12 M^4}\langle H(^3P_2) | \mathcal P_{8a\,\textrm{cm}} |H(^3P_2) \rangle    
\nn\\
&&
+ \frac{\als^2 \pi}{M^5}\left(\frac{1}{5} + \frac{25}{864}n_f\right)
  \langle H(^3P_2) | \mathcal T_{1\textrm{-}8} (^3P_2,^3S_1)|H(^3P_2) \rangle 
+ \frac{1}{2}\frac{ \als^2 \pi}{ M^4} \langle H(^3P_2) | \mathcal O_8 (^1P_1) |H(^3P_2)\rangle 
\nn\\
&&
+ \frac{5}{6}\frac{\als^2 \pi}{ M^2}\langle H(^3P_2) |\mathcal O_8 (^1S_0) |H(^3P_2) \rangle
+ \frac{4}{9} \frac{\als^2 \pi}{M^2}\langle H(^3P_2) |\mathcal O_1 (^1S_0) |H(^3P_2) \rangle   
\nn\\
&&
+ \frac{2}{3} \frac{\als^2 \pi}{M^4} \langle H(^3P_2) | \mathcal O_8 (^3P_2) |H(^3P_2) \rangle 
+ \frac{ \als^2 \pi}{ M^6}\left( - \frac{2}{3} +  \frac{23}{54} n_f \right)
  \langle H(^3P_2) | \mathcal Q'_8(^3S_1) | H(^3P_2) \rangle 
\nn \\
&& 
+\frac{ \als^2 \pi}{ M^6}\left(- \frac{1}{30}  + \frac{5}{9} n_f \right)
\langle H(^3P_2) | \mathcal Q'_8(^3S_1,^3D_1) | H(^3P_2) \rangle 
\nn \\
&& 
+\frac{ \als^2 \pi}{ M^6}\left(\frac{1}{2} + \frac{1}{12} n_f \right)
\langle H(^3P_2) | \mathcal Q_8(^3D_1) | H(^3P_2) \rangle 
\nn \\
&& 
+ \frac{1}{5} \frac{\als^2 \pi}{M^6}\langle H(^3P_2) | \mathcal Q_8(^3D_2) | H(^3P_2) \rangle 
+ \frac{2}{7} \frac{\als^2 \pi}{M^6}\langle H(^3P_2) | \mathcal Q_8(^3D_3) | H(^3P_2) \rangle 
\nn \\
&& 
-\frac{80}{189}\frac{\als^2 \pi}{M^6}\langle H(^3P_2) | \mathcal P_1(^3P_2, ^3F_2) | H(^3P_2) \rangle
-\frac{50}{63}\frac{\als^2 \pi}{M^6}\langle H(^3P_2) | \mathcal P_8(^3P_2, ^3F_2) | H(^3P_2) \rangle
\,.
\label{eq:3p2width}
\eea

A general source of concern is the proliferation of matrix elements 
with the increasing order of the expansion  in $v$. Spin symmetry and vacuum saturation 
\cite{Bodwin:1994jh} may help to reduce the number of matrix elements by 
relating different spin states and hadronic with electromagnetic matrix elements. 
The actual number of independent matrix elements depends on the power counting.

In the power counting of \cite{Bodwin:1994jh}, only the first six matrix elements of 
Eqs. \eqref{eq:3p0width} and \eqref{eq:3p2width} and the first five of 
\eqref{eq:3p1width} contribute. In \cite{Huang:1997nt}, it was assumed that only 
the first four matrix elements of Eqs. \eqref{eq:3p0width} and \eqref{eq:3p2width} and the first three of 
\eqref{eq:3p1width} contribute.

The conservative power counting adopted here has been suggested in \cite{Brambilla:2001xy} to be appropriate 
when $\lQ \gg mv^2$. Under this condition, matrix elements are non-perturbative quantities 
and should be evaluated on the lattice. One can also take advantage of the factorization provided by potential NRQCD  
\cite{Brambilla:1999xf,Brambilla:2001xy,Brambilla:2003mu}.
According to it, the matrix elements can be factorized into the product of the quarkonium wave function
in the origin squared (or derivatives of it)  and few universal non-perturbative
correlation function, eventually achieving a reduction in the number and a simplification 
of the non-perturbative operators needed.

We also note that the convergence of the perturbative series of the matching 
coefficient is typically poor. For a discussion and references 
we refer for instance to \cite{Vairo:2002iw}.

Phenomenological applications of the expressions of the decay widths 
will therefore entail work in two  complimentary directions: 
(1) improving the knowledge of the NRQCD matrix elements 
either by direct evaluation, for example by fitting the experimental data, by lattice calculations, 
and by models, or by exploiting the hierarchy of scales still entangled in NRQCD using 
EFTs of lower energy, like potential NRQCD; (2) improving the convergence of the perturbative 
series of the matching coefficients by resumming large contributions either 
related to large logarithms, or of the type discussed, for instance, in~\cite{Bodwin:2001pt}.

\acknowledgments
Part of this work has been carried out at the IFIC, Valencia. N.B.
and A.V. gratefully acknowledge the warm hospitality of the IFIC members.
N.B. and A.V.  acknowledge financial support from ``Azioni Integrate 
Italia-Spagna 2004 (IT1824)/Acciones Integradas  Espa\~na-Italia (HI2003-0362)'', 
and from the cooperation agreement INFN05-04 (MEC-INFN) and the European 
Research Training Network FLAVIA{\it net} (FP6, Marie Curie Programs, Contract MRTN-CT-2006-035482).
E.M. acknowledges support by the U.S. Department of Energy under grant number DE-FG02-06ER41449.

\appendix
\section{Summary and definition of the NRQCD operators}
\label{AppA}
The two-fermion sector of the NRQCD Lagrangian relevant for the matching discussed in section \ref{matching}
is:
\bea
{\cal L}_{\textrm{2-f}}  &=&
\psi^\dagger \left(i D_0 + \frac{{\vec D}^2}{2M}
+ \frac{{\vec \sigma} \cdot g {\vec B}}{2 M}
+ \frac{({\vec D}\cdot g{\vec E})}{8 M^2}
- \frac{{\vec \sigma} \cdot [-i {\vec D}\times, g{\vec E}]}{8 M^2}
+ \frac{({\vec D}^2)^2}{8M^3}
+ \frac{\{ {\vec D}^2, {\vec \sigma} \cdot g {\vec B} \}}{8 M^3}
\right.
\nn\\
&&
\left.
- \frac{3}{64 M^4} \{ {\vec D}^2, {\vec \sigma} \cdot [-i {\vec D}\times,
  g{\vec E}] \}
+ \frac{3}{64 M^4} \{ {\vec D}^2, ({\vec D}\cdot g{\vec E}) \}
+ \frac{{\vec D}^6}{16M^5}  \right) \psi
\nn\\
&&
+ ~ \textrm{c.c.}\,,
\label{NRQCD:bilinear}
\eea
where $\sigma^i$ are the Pauli matrices,
$i D_0 = i \partial_0 - t^a \, g A^a_0$, $i {\vec D} = i \vec\nabla + t^a \, g {\vec A}^a$,
$[{\vec D} \times, {\vec E}]={\vec D} \times {\vec E} - {\vec E} \times {\vec D}$,
${E}^i = F^{i0}$ and ${B}^i = -\epsilon_{ijk}F^{jk}/2$ ($\epsilon_{123} = 1$).
We have not displayed terms of order $1/M^6$ or smaller and matching
coefficients of ${O}(\als)$ or smaller. The general structure of the four-fermion sector of the NRQCD
Lagrangian is: 
\be 
{\cal L}_{\textrm{4-f}}  = \sum_n \frac{c^{(n)}}{{M^{d_n - 4}}} \mathcal O^{(n)}_{\textrm{4-f}}. 
\ee 
Here, we list the operators relevant for the matching performed in section
\ref{matching} ordered by dimension. We use  $\overleftrightarrow D \equiv
\overrightarrow D - \overleftarrow D$. 

For the octet operators defined in Eqs. \eqref{eq:B.2oct}, 
\eqref{eq:B.3oct}, \eqref{eq:B.11} and \eqref{eq:B.4oct}-\eqref{eq:B.6oct}, since the covariant 
derivative $\overleftrightarrow D$ does not commute with the color matrix $t^a$
we need to specify the ordering between the two and verify that the resulting operator is gauge invariant. 
Let us consider, for example, the operator $\mathcal O_8(^1P_1)$:
\begin{equation}
\label{eq:gauge.1}
\mathcal O_8(^1P_1) = \psi^{\dag} \overleftrightarrow D t^a \chi
\chi^{\dag} \overleftrightarrow D t^a \psi,
\end{equation}
and the three different orderings:
\begin{eqnarray}
\left[ \psi^{\dag} \overleftrightarrow D t^a \chi\right]^{(1)} & \equiv -\left(
\vec D \psi \right)^{\dag} t^a \chi  + \psi^{\dag}\vec D t^a \chi\,,
\label{eq:gauge.2.1} \\
\left[\psi^{\dag} \overleftrightarrow D t^a \chi \right]^{(2)} & \equiv
-\left(\vec D t^a \psi\right)^{\dag}\chi  + \psi^{\dag} t^a \vec D \chi \,,
\label{eq:gauge.2.2} \\
\left[\psi^{\dag} \overleftrightarrow D t^a \chi \right]^{(3)} & \equiv
-\left(\vec D  \psi\right)^{\dag}t^a\chi  + \psi^{\dag} t^a \vec D \chi \,.
\label{eq:gauge.2.3bis}
\end{eqnarray}
Under the gauge transformation
\begin{equation}
\label{eq:gauge.3}
\begin{split}
& \psi \rightarrow \left( 1 + i \omega^a t^a  \right)\psi,
\\
& \chi \rightarrow \left( 1 + i \omega^a t^a  \right)\chi,
\\
&  A^{a\mu} \rightarrow A^{a\mu} - \frac{1}{g} \partial^{\mu} \omega^a + f^{abc}A^{b\mu}\omega^c,
\end{split}
\end{equation}
\eqref{eq:gauge.2.1}, \eqref{eq:gauge.2.2} and \eqref{eq:gauge.2.3bis} transform respectively as:
\begin{eqnarray}
\delta \left[ \psi^{\dag} \overleftrightarrow D t^a \chi \right]^{(1)} = &
 f^{abc} \omega^c \left[ \psi^{\dag} \overleftrightarrow D t^b \chi  \right]^{(1)} 
+ f^{abc} \vec{\partial} \, \omega^c \psi^{\dag} t^b \chi,
\label{eq:gauge.3.1}\\
\delta \left[ \psi^{\dag} \overleftrightarrow D t^a \chi \right]^{(2)} = &
 f^{abc} \omega^c \left[ \psi^{\dag}  \overleftrightarrow D  t^b  \chi  \right]^{(2)} 
- f^{abc} \vec{\partial} \, \omega^c \psi^{\dag} t^b \chi,
\label{eq:gauge.3.2}\\ 
\delta \left[ \psi^{\dag} \overleftrightarrow D   t^a \chi \right]^{(3)} = &
 f^{ a bc}\omega^c \left[ \psi^{\dag} \overleftrightarrow D t^b \chi \right]^{(3)}.
\label{eq:gauge.3.3}
\end{eqnarray}
Only the last ordering leads to a gauge invariant definition of $\mathcal O_8(^1P_1)$:
\begin{equation}
\label{eq:gauge.4} 
\delta \, \mathcal O_8(^1P_1) =
 f^{ a bc}\omega^c \left( 
\left[ \psi^{\dag} \overleftrightarrow D  t^b \chi \right]^{(3)}
\left[ \chi^{\dag}  \overleftrightarrow D t^a\psi \right]^{(3)}
+ \left[\psi^{\dag} \overleftrightarrow D  t^a \chi \right]^{(3)}
  \left[\chi^{\dag}  \overleftrightarrow D  t^b\psi \right]^{(3)} \right) = 0\,.
\end{equation}
Therefore, we define
\begin{equation}
\label{eq:gauge.2.3}
\psi^{\dag} \overleftrightarrow D t^a \chi  \equiv
-\left(\vec D  \psi\right)^{\dag}t^a\chi  + \psi^{\dag} t^a \vec D
\chi.
\end{equation}
Generalizing to operators containing more than one covariant derivative, we define 
\begin{equation}\label{eq:gauge.6}
\begin{gathered}
\psi^{\dag} \overleftrightarrow D^{i_1} \ldots \overleftrightarrow
D^{i_n} t^a \chi \equiv (-1)^{n} \left(\vec D^{i_1} \ldots \vec D^{i_n}
\psi \right)^{\dag} t^a \chi  \\+ (-1)^{n-1} \left(\vec D^{i_2}
\ldots \vec D^{i_n} \psi \right)^{\dag} t^a \vec D^{i_1} \chi +
\ldots +
  \psi^{\dag}
t^a \vec D^{i_1} \ldots \vec D^{i_n}\chi.
\end{gathered}
\end{equation}

The singlet-octet transition operators are denoted by $\mathcal O_{1\textrm{-}8}(^{2S+1}L_J, ^{2S'+1}L'_{J'})$  or 
$\mathcal O_{8\textrm{-}1}$ $(^{2S+1}L_J, ^{2S'+1}L'_{J'})$.
In the first case the first set of quantum numbers refers to the $Q \bar Q$ pair in a color singlet state, the second
to the $Q \bar Q$ pair in the octet state, while in the second 
case the first set of quantum numbers refers to the $Q \bar Q$ 
pair in a color octet state and  the second one
to the $Q \bar Q$ pair in the singlet state.
In some cases, it has been found convenient to introduce singlet-octet transition operators 
that annihilate (create) states containing a $Q \bar Q$ pair and a gluon in 
which the total angular momentum $J$ of the the quark-antiquark pair has not a definite value
(a definite value could be attributed by further decomposing these operators in irreducible 
spherical tensors). 
This is the case of the operators $\mathcal T^{(i)}_{1\textrm{-}8}(^3S_1,^3P)$ 
and $\mathcal T^{(i)}_{8\textrm{-}1}(^3S_1,^3P)$. In these cases, 
we cannot use the quantum number $J$ and we have to denote the state just by the orbital 
angular momentum and spin quantum numbers.  

The symbols $A^{(i} B^{j)}$ and $S^{((ij)}A^{k)}$, used in the definitions of some four-fermion operators
denote symmetric and traceless two and three indices tensors, according to
\be
A^{(i}B^{j)}=\frac{A^iB^j+A^jB^i}{2}-\frac{\delta^{ij}}{3}\vec A
\cdot \vec B,
\label{eq:B.b}
\ee

\be
S^{((ij)}A^{k)} =
\frac{1}{3}\left(S^{(ij)}A^k+S^{(ik)}A^j+S^{(jk)}A^i\right)
-\frac{2}{15}\left(\delta^{ij}\delta^{lk}+\delta^{ik}\delta^{lj}+\delta^{jk}\delta^{li}
\right)S^{(ml)}A^m.
\label{eq:B.d}
\ee
For some details on the decomposition of Cartesian tensors in terms of irreducible 
spherical tensors see \cite{Brambilla:2006ph}.

\paragraph{Operators of dimension 6}
\be
\label{eq:B.1}
\begin{split}
\mathcal{O}_{1}(^1S_0) = & \psi^{\dag} \chi \, \chi^{\dag}\psi ,
\\
\mathcal O_{1}(^3S_1)= & \psi^{\dag} \vec{\sigma} \chi  \cdot
\chi^{\dag} \vec{\sigma} \psi.
\end{split}
\ee

\be
\label{eq:B.1oct}
\begin{split}
\mathcal{O}_{8}(^1S_0) = & \psi^{\dag} t^a \chi \, \chi^{\dag} t^a\psi,
\\
\mathcal O_{8}(^3S_1)= & \psi^{\dag}  \vec{\sigma} \, t^a \chi  \cdot
\chi^{\dag}  \vec{\sigma} \, t^a \psi.
\end{split}
\ee

\paragraph{Operators of dimension 8}
\be
\label{eq:B.2}
\begin{split}
\mathcal{P}_{1}(^1S_0) = & \frac{1}{2}
\psi^{\dag}\left(-\frac{i}{2}\overleftrightarrow D \right)^2\chi
\, \chi^{\dag}\psi + \textrm{H.c.}\,,
\\
\mathcal{P}_{1}(^3S_1)= & \frac{1}{2} \psi^{\dag} \vec{\sigma}
\chi \cdot \chi^{\dag} \vec{\sigma}
\left(-\frac{i}{2}\overleftrightarrow D\right)^2\psi
+\textrm{H.c.}\,,
\\
\mathcal P_{1}(^3S_1,^3D_1)= & \frac{1}{2} \psi^{\dag} \sigma^i
\chi \,  \chi^{\dag} \sigma^j \left(-\frac{i}{2}\right)^2
\overleftrightarrow D^{(i} \overleftrightarrow D^{j)}\psi
+\textrm{H.c.}\,.
\end{split}
\ee

\be
\label{eq:B.3}
\begin{split}
\mathcal{O}_{1}(^1P_1) = &
\psi^{\dag}\left(-\frac{i}{2}\overleftrightarrow D \right)\chi \cdot
\chi^{\dag}\left(-\frac{i}{2}\overleftrightarrow D \right)\psi,
\\
\mathcal O_{1}(^3P_0)= & \frac{1}{3}  \psi^{\dag}
\left(-\frac{i}{2}\overleftrightarrow{D}\cdot\vec{\sigma}\right)
\chi \,  \chi^{\dag}
\left(-\frac{i}{2}\overleftrightarrow{D}\cdot\vec{\sigma}\right)\psi,
\\
\mathcal O_{1}(^3P_1)= & \frac{1}{2}
\psi^{\dag}\left(-\frac{i}{2}\overleftrightarrow{D}\times\vec{\sigma}\right)
\chi \cdot  \chi^{\dag}
\left(-\frac{i}{2}\overleftrightarrow{D}\times\vec{\sigma}\right)
\psi,
\\
\mathcal O_{1}(^3P_2)= &
\psi^{\dag}\left(-\frac{i}{2}\overleftrightarrow{D}^{(i}\sigma^{j)}\right)\chi
\, \chi^{\dag}
\left(-\frac{i}{2}\overleftrightarrow{D}^{(i}\sigma^{j)}\right)
\psi.
\end{split}
\ee

\be
\label{eq:B.2oct}
\begin{split}
\mathcal{P}_{8}(^1S_0) = & \frac{1}{2}
\psi^{\dag}  \left(-\frac{i}{2}\overleftrightarrow D \right)^2 t^a\chi
\, \chi^{\dag} t^a\psi + \textrm{H.c.}\,,
\\
\mathcal{P}_{8}(^3S_1)= & \frac{1}{2} \psi^{\dag}  \vec{\sigma}
\, t^a \chi \cdot \chi^{\dag}  \vec{\sigma}
\left(-\frac{i}{2}\overleftrightarrow D\right)^2 t^a\psi
+\textrm{H.c.}\,,
\\
\mathcal P_{8}(^3S_1,^3D_1)= & \frac{1}{2} \psi^{\dag}  \sigma^i
\, t^a \chi \,  \chi^{\dag}  \sigma^j \left(-\frac{i}{2}\right)^2
\overleftrightarrow D^{(i} \overleftrightarrow D^{j)} \, t^a \psi
+\textrm{H.c.}\,.
\end{split}
\ee

\be
\label{eq:B.3oct}
\begin{split}
\mathcal{O}_{8}(^1P_1) = &
\psi^{\dag} \left(-\frac{i}{2}\overleftrightarrow D \right)t^a \chi \cdot
\chi^{\dag}\left(-\frac{i}{2}\overleftrightarrow D \right)t^a\psi,
\\
\mathcal O_{8}(^3P_0)= & \frac{1}{3}  \psi^{\dag}
\left(-\frac{i}{2}\overleftrightarrow{D}\cdot\vec{\sigma}\right)
t^a \chi \,  \chi^{\dag}
\left(-\frac{i}{2}\overleftrightarrow{D}\cdot\vec{\sigma}\right) t^a \psi,
\\
\mathcal O_{8}(^3P_1)= & \frac{1}{2}
\psi^{\dag}\left(-\frac{i}{2}\overleftrightarrow{D}\times\vec{\sigma}\right)
t^a \chi \cdot  \chi^{\dag}
\left(-\frac{i}{2}\overleftrightarrow{D}\times\vec{\sigma}\right)
t^a \psi,
\\
\mathcal O_{8}(^3P_2)= &
\psi^{\dag}\left(-\frac{i}{2}\overleftrightarrow{D}^{(i}\sigma^{j)}\right) t^a \chi
\, \chi^{\dag}
\left(-\frac{i}{2}\overleftrightarrow{D}^{(i}\sigma^{j)}\right)
t^a \psi.
\end{split}
\ee

\begin{equation}
\label{eq:B.11}
\begin{split}
\mathcal O_{1\,\textrm{cm}} & =  \psi^{\dag} \left( \Dlr \right)
\times \vec{\sigma} \chi \, \cdot
          \vec{\nabla} \left( \chi^{\dag} \psi \right) + \textrm{H.c.}  \,,
\\
\mathcal O'_{1\,\textrm{cm}} & =   -\psi^{\dag} \left( \Dlr
\right) \chi \cdot
        \,  \vec{\nabla} \times \left( \chi^{\dag} \vec{\sigma} \psi \right) + \textrm{H.c.}\,,  
\\
\mathcal O_{8\,\textrm{cm}} & =   \psi^{\dag}  \left( \Dlr
\right)\times \vec{\sigma} t^a \chi \, \cdot
          \vec{D}_{ab} \left( \chi^{\dag} t^b \psi \right) +\textrm{H.c.} \,, 
\\
\mathcal O'_{8\,\textrm{cm}} & = -  \psi^{\dag} \left( \Dlr
\right) t^a \chi \,
       \cdot   \vec{D}_{ab} \times \left( \chi^{\dag}t^b \vec \sigma\psi \right) + \textrm{H.c.}\,.
\end{split}
\end{equation}

\begin{equation}
\label{eq:B.12}
\begin{split}
\mathcal P_{1a\,\textrm{cm}} & = \nabla^i  \left(\psi^{\dag}
\sigma^j \chi\right) \, \nabla^i  \left( \chi^{\dag} \sigma^j
\psi\right)\,,
\\
\mathcal P_{1b\,\textrm{cm}} & = \vec{\nabla} \cdot
\left(\psi^{\dag} \vec{\sigma} \chi\right) \, \vec{\nabla} \cdot
\left( \chi^{\dag} \vec{\sigma} \psi\right) \,,
\\
\mathcal P_{1c\,\textrm{cm}} & = \vec{\nabla} \left(\psi^{\dag}
 \chi\right) \, \cdot \vec{\nabla}  \left( \chi^{\dag} \psi\right) \,,
\\
 \mathcal P_{8a\,\textrm{cm}} & = D^i_{ab}  \left(\psi^{\dag}
t^a \sigma^j \chi\right) \, D^i_{ac}  \left( \chi^{\dag} t^c
\sigma^j \psi\right)\,,
\\
\mathcal P_{8b\,\textrm{cm}} & = \vec{D}_{ab} \cdot
\left(\psi^{\dag} t^b\vec{\sigma} \chi\right) \, \vec{D}_{ac}
\cdot
\left( \chi^{\dag} t^c \vec{\sigma} \psi\right) \,,
\\
\mathcal P_{8c\,\textrm{cm}} & = \vec{D}_{ab} \left(\psi^{\dag}
 t^b \chi\right) \, \cdot \vec{D}_{ac}  \left( \chi^{\dag} t^c \psi\right)\,.
\end{split}
\end{equation}

\be
\label{eq:B.8}
\begin{split}
\mathcal S_{1\textrm{-}8}(^1S_0,^3S_1) = & \frac{1}{2} \psi^{\dag} g \vec B \cdot
{\vec\sigma} \chi \,  \chi^{\dag} \psi + \textrm{H.c.}\,,
\\
\mathcal S_{1\textrm{-}8}(^3S_1,^1S_0) = & \frac{1}{2} \psi^{\dag} g \vec B \chi \cdot
\chi^{\dag}\vec{\sigma}\psi + \textrm{H.c.}\,.
\end{split}
\ee

\paragraph{Operators of dimension 9}
\be
\label{eq:B.9}
\begin{split}
\mathcal T_{1\textrm{-}8}(^1S_0,^1P_1) =& \frac{1}{2} \psi^{\dag}\chi \,
\chi^{\dag} (\overleftrightarrow D \cdot  g\vec{E}+ g\vec E \cdot
\overleftrightarrow{D} )\psi + \textrm{H.c.}\,,
\\
\mathcal F_{8\textrm{-}8}(^1S_0,^1P_1) =& \frac{1}{2} f^{abc}\, \psi^{\dag}t^a\chi
\, \chi^{\dag}t^b (\overleftrightarrow D \cdot  g\vec{E^c}+ g\vec
E^c \cdot \overleftrightarrow{D} )\psi + \textrm{H.c.}\,,
\\
\mathcal D_{8\textrm{-}8}(^1S_0,^1P_1) =& \frac{1}{2} d^{abc}\, \psi^{\dag}t^a \chi
\, \chi^{\dag} t^b (\overleftrightarrow D \cdot  g\vec{E^c}+ g\vec
E^c \cdot \overleftrightarrow{D} )\psi + \textrm{H.c.}\,,
\\
\mathcal D^{(0)}_{8\textrm{-}8}(^3S_1,^3P) =& \frac{1}{6} d^{abc}
\psi^{\dag}t^a\vec{\sigma}\chi \cdot \chi^{\dag} \vec{\sigma}
\, (\overleftrightarrow D \cdot  g\vec{E^b}+ g\vec E^b \cdot
\overleftrightarrow{D} )t^c \psi + \textrm{H.c.}\,,
\\
\mathcal D^{(2)}_{8\textrm{-}8}(^3S_1,^3P) =& \frac{1}{2} d^{abc}
\psi^{\dag} t^a \sigma^i\chi \chi^{\dag} \sigma^j (\overleftrightarrow
D^{(i}  g \vec{E}^{b j)}+ g \vec E^{(b i}\overleftrightarrow{D}^{j)})t^c\psi +\textrm{H.c.}\,,
\\
\mathcal F_{8}(^1S_0) =& \frac{i}{2} f^{abc}\, \psi^{\dag}
(\vec D \cdot \vec E)^b\,t^c\chi \, \chi^{\dag}t^a \psi + \textrm{H.c.}\,,
\\
\mathcal T_{1\textrm{-}8}(^1P_1,^1S_0) =& \frac{1}{2} \psi^{\dag} g\vec E \chi \cdot
\chi^{\dag} \overleftrightarrow D \psi + \textrm{H.c.}\,,
\\
\mathcal T^{(0)}_{1\textrm{-}8}(^3S_1,^3P) =& \frac{1}{6}
\psi^{\dag}\vec{\sigma}\chi \cdot \chi^{\dag} \vec{\sigma}
(\overleftrightarrow D \cdot  g\vec{E}+ g\vec E \cdot
\overleftrightarrow{D} )\psi + \textrm{H.c.}\,,
\\
\mathcal T^{(1)}_{1\textrm{-}8}(^3S_1,^3P) =& \frac{1}{4}
\psi^{\dag}\vec{\sigma}\chi \cdot \chi^{\dag} \vec{\sigma} \times
(-\overleftrightarrow D \times  g \vec{E} -  g \vec E \times
\overleftrightarrow{D})\psi + \textrm{H.c.}\,,
\\
\mathcal T^{(1)\prime}_{1\textrm{-}8}(^3S_1,^3P) =& \frac{1}{4}
\psi^{\dag}\vec{\sigma}\chi \cdot \chi^{\dag} \vec{\sigma} \times
(\overleftrightarrow D \times  g \vec{E} -  g \vec E \times
\overleftrightarrow{D})\psi + \textrm{H.c.}\,,
\\
\mathcal T^{(1)\prime}_{8\textrm{-}1}(^3S_1,^3P) =& \frac{1}{4}
\psi^{\dag}t^a\vec{\sigma}\chi \cdot \chi^{\dag} \vec{\sigma} \times
(\overleftrightarrow D \times  g \vec{E^a} -  g \vec E^a \times
\overleftrightarrow{D})\psi + \textrm{H.c.}\,,
\\
\mathcal T^{(2)}_{1\textrm{-}8}(^3S_1,^3P) =& \frac{1}{2} \psi^{\dag}\sigma^i\chi
\, \chi^{\dag} \sigma^j (\overleftrightarrow D^{(i}  g
\vec{E}^{j)}+ g \vec E^{(i} \overleftrightarrow{D}^{j)})\psi
+\textrm{H.c.}\,,
\\
\mathcal T_{1\textrm{-}8}(^3P_0,^3S_1) =& \frac{1}{6} \psi^{\dag}
\left(\overleftrightarrow D \cdot \vec{\sigma}\right) \chi \,
\chi^{\dag} \vec{\sigma}\cdot g \vec{E} \psi +\textrm{H.c.}\,,
\\
\mathcal T_{1\textrm{-}8}(^3P_1,^3S_1) =& \frac{1}{4} \psi^{\dag}
\left(\overleftrightarrow D \times \vec{\sigma}\right) \chi \cdot
\chi^{\dag} \vec{\sigma}\times g \vec{E} \psi +\textrm{H.c.}\,,
\\
\mathcal T_{1\textrm{-}8}(^3P_2,^3S_1) =& \frac{1}{2} \psi^{\dag}
\left(\overleftrightarrow  D^{(i} \sigma^{j)}\right) \chi \,
\chi^{\dag} \sigma^{(i} gE^{j)} \psi +\textrm{H.c.}\,.
\end{split}
\ee

\paragraph{Operators of dimension 10}
\be
\label{eq:B.4}
\begin{split}
\mathcal{Q'}_{1}(^1S_0) = &
\psi^{\dag}\left(-\frac{i}{2}\overleftrightarrow D \right)^2 \chi
\, \chi^{\dag}\left(-\frac{i}{2}\overleftrightarrow D
\right)^2\psi,
\\
\mathcal{Q''}_{1}(^1S_0) = & \frac{1}{2}
\psi^{\dag}\left(-\frac{i}{2}\overleftrightarrow D \right)^4\chi
\, \chi^{\dag}\psi + \textrm{H.c.}\,,
\\
\mathcal Q'_{1}(^3S_1)= &
\psi^{\dag}\left(-\frac{i}{2}\overleftrightarrow D \right)^2
\vec{\sigma}\chi \cdot \chi^{\dag}
\left(-\frac{i}{2}\overleftrightarrow D \right)^2
\vec{\sigma}\psi,
\\
\mathcal Q_{1}''(^3S_1)= & \frac{1}{2} \psi^{\dag}
\left(-\frac{i}{2}\overleftrightarrow D \right)^4 \vec{\sigma}
\chi \cdot  \chi^{\dag}\vec{\sigma}\psi +\textrm{H.c.}\,,
\\
\mathcal Q_{1}'(^3S_1,^3D_1)= & \frac{1}{2} \psi^{\dag}
\left(-\frac{i}{2}\right)^2 \overleftrightarrow D^{(i}
\overleftrightarrow D^{j)}\sigma^i\chi \, \chi^{\dag}\sigma^j
\left(-\frac{i}{2}\overleftrightarrow D \right)^2 \psi
+\textrm{H.c.}\,,
\\
\mathcal Q_{1}''(^3S_1,^3D_1)= & \frac{1}{2}
\psi^{\dag}\left(-\frac{i}{2}\overleftrightarrow D \right)^2
\left(-\frac{i}{2}\right)^2 \overleftrightarrow D^{(i}
\overleftrightarrow D^{j)}\sigma^i\chi \,  \chi^{\dag}\sigma^j\psi
+\textrm{H.c.}\,.
\end{split}
\ee

\be
\label{eq:B.5}
\begin{split}
\mathcal{P}_{1}(^1P_1) = & \frac{1}{2}
\psi^{\dag}\left(-\frac{i}{2}\overleftrightarrow D
\right)^2\left(-\frac{i}{2} \overleftrightarrow D^i \right)\chi \,
\chi^{\dag}\left(-\frac{i}{2} \overleftrightarrow D^i\right)\psi +
\textrm{H.c.}\,,
\\
\mathcal P_{1}(^3P_0)= & \frac{1}{6} \psi^{\dag} \left(
-\frac{i}{2}\overleftrightarrow D \cdot \vec{\sigma} \right)
\left(-\frac{i}{2}\overleftrightarrow D\right)^2 \chi \,
\chi^{\dag}\left(-\frac{i}{2}\overleftrightarrow D \cdot
\vec{\sigma}\right)\psi + \textrm{H.c.}\,,
\\
\mathcal P_{1}(^3P_1)= & \frac{1}{4} \psi^{\dag} \left(
-\frac{i}{2}\overleftrightarrow D \times \vec{\sigma} \right)
\left(-\frac{i}{2}\overleftrightarrow D\right)^2 \chi \cdot
\chi^{\dag}\left(-\frac{i}{2}\overleftrightarrow D \times
\vec{\sigma}\right)\psi + \textrm{H.c.}\,,
\\
\mathcal P_{1}(^3P_2)= & \frac{1}{2}
\psi^{\dag}\left(-\frac{i}{2}\overleftrightarrow D^{(i}
\sigma^{j)}\right) \left(- \frac{i}{2} \overleftrightarrow  D \right)^2
\chi \,  \chi^{\dag}\left(-\frac{i}{2}\overleftrightarrow
D^{(i}\sigma^{j)}\right)\psi + \textrm{H.c.}\,,
\\
\mathcal P_{1}(^3P_2,^3F_2)= & \frac{1}{2}
\psi^{\dag}\left(-\frac{i}{2}\right)^2\overleftrightarrow
D^{(i}\overleftrightarrow D^{j)}
\left(-\frac{i}{2}\overleftrightarrow D \cdot\vec{\sigma}\right)
\chi \, \chi^{\dag}\left(-\frac{i}{2}\overleftrightarrow
D^{(i}\sigma^{j)}\right)\psi
\\
& - \frac{1}{5}
\psi^{\dag}\left(-\frac{i}{2}\right)\overleftrightarrow D^{(i}
\sigma^{j)} \left(- \frac{i}{2} \overleftrightarrow  D \right)^2
\chi \, \chi^{\dag}\left(-\frac{i}{2}\overleftrightarrow
D^{(i}\sigma^{j)}\right)\psi + \textrm{H.c.}\,.
\end{split}
\ee
\be
\label{eq:B.6}
\begin{split}
\mathcal{Q}_{1}(^1D_2) = &
\psi^{\dag}\left(-\frac{i}{2}\right)^2\overleftrightarrow D^{(i}
\overleftrightarrow D^{j)}\chi \,
\chi^{\dag}\left(-\frac{i}{2}\right)^2\overleftrightarrow D^{(i}
\overleftrightarrow D^{j)}\psi,
\\
\mathcal Q_{1}(^3D_3)= & \psi^{\dag}\left(-\frac{i}{2}\right)^2
\overleftrightarrow D^{((i} \overleftrightarrow D^{j)}
\sigma^{l)}\chi \, \chi^{\dag}\left(-\frac{i}{2}\right)^2
\overleftrightarrow D^{((i} \overleftrightarrow D^{j)}
\sigma^{l))}\psi,
\\
\mathcal Q_{1}(^3D_2)= &
\frac{2}{3}
\psi^{\dag}
\left(-\frac{i}{2}\right)^2 \left(\varepsilon^{ilm}\overleftrightarrow D^{(j}  \overleftrightarrow D^{l)} \sigma^{m}
+ \frac{1}{2}\varepsilon^{ijl}\overleftrightarrow D^{(m}  \overleftrightarrow D^{l)} \sigma^{m}\right) \chi
\\
& \times
 \chi^{\dag} \left(-\frac{i}{2}\right)^2 
\left(\varepsilon^{inp}\overleftrightarrow D^{(j}  \overleftrightarrow D^{n)} \sigma^{p}
+ \frac{1}{2}\varepsilon^{ijn}\overleftrightarrow D^{(p}  \overleftrightarrow D^{n)} \sigma^{p}\right) \psi,
\\
\mathcal Q_{1} (^3D_1) = & \psi^{\dag}\left(-\frac{i}{2}\right)^2
\overleftrightarrow D^{(i} \overleftrightarrow D^{j)}
\sigma^{i}\chi \,  \chi^{\dag}\left(-\frac{i}{2}\right)^2
\overleftrightarrow D^{(l} \overleftrightarrow D^{j)}
\sigma^{l}\psi.
\end{split}
\ee

\be
\label{eq:B.4oct}
\begin{split}
\mathcal{Q'}_{8}(^1S_0) = &
\psi^{\dag}\left(-\frac{i}{2}\overleftrightarrow D \right)^2 t^a \chi
\, \chi^{\dag}\left(-\frac{i}{2}\overleftrightarrow D
\right)^2 t^a\psi,
\\
\mathcal{Q''}_{8}(^1S_0) = & \frac{1}{2}
\psi^{\dag}\left(-\frac{i}{2}\overleftrightarrow D \right)^4 t^a\chi
\, \chi^{\dag} t^a \psi + \textrm{H.c.}\,,
\\
\mathcal Q'_{8}(^3S_1)= &
\psi^{\dag}\left(-\frac{i}{2}\overleftrightarrow D \right)^2
\vec{\sigma} \, t^a \chi \cdot \chi^{\dag}
\left(-\frac{i}{2}\overleftrightarrow D \right)^2
\vec{\sigma}\, t^a \psi,
\\
\mathcal Q_{8}''(^3S_1)= & \frac{1}{2} \psi^{\dag}
\left(-\frac{i}{2}\overleftrightarrow D \right)^4 \vec{\sigma} \, t^a
\chi \cdot  \chi^{\dag}\vec{\sigma} \, t^a\psi +\textrm{H.c.}\,,
\\
\mathcal Q_{8}'(^3S_1,^3D_1)= & \frac{1}{2} \psi^{\dag}
\left(-\frac{i}{2}\right)^2 \overleftrightarrow D^{(i}
\overleftrightarrow D^{j)}\sigma^i \, t^a \chi \, \chi^{\dag}\sigma^j
\left(-\frac{i}{2}\overleftrightarrow D \right)^2 t^a  \psi
+\textrm{H.c.}\,,
\\
\mathcal Q_{8}''(^3S_1,^3D_1)= & \frac{1}{2}
\psi^{\dag}\left(-\frac{i}{2}\overleftrightarrow D \right)^2
\left(-\frac{i}{2}\right)^2 \overleftrightarrow D^{(i}
\overleftrightarrow D^{j)}\sigma^i \, t^a \chi \,  \chi^{\dag}\sigma^j t^a \,\psi
+\textrm{H.c.}\,.
\end{split}
\ee

\be
\label{eq:B.5oct}
\begin{split}
\mathcal{P}_{8}(^1P_1) = & \frac{1}{2}
\psi^{\dag}\left(-\frac{i}{2}\overleftrightarrow D
\right)^2\left(-\frac{i}{2} \overleftrightarrow D^i \right) t^a \chi \,
\chi^{\dag}\left(-\frac{i}{2} \overleftrightarrow D^i\right) t^a \psi +
\textrm{H.c.}\,,
\\
\mathcal P_{8}(^3P_0)= & \frac{1}{6} \psi^{\dag} \left(
-\frac{i}{2}\overleftrightarrow D \cdot \vec{\sigma} \right)
\left(-\frac{i}{2}\overleftrightarrow D\right)^2 t^a \chi \,
\chi^{\dag}\left(-\frac{i}{2}\overleftrightarrow D \cdot
\vec{\sigma}\right) t^a \psi + \textrm{H.c.}\,,
\\
\mathcal P_{8}(^3P_1)= & \frac{1}{4} \psi^{\dag} \left(
-\frac{i}{2}\overleftrightarrow D \times \vec{\sigma} \right)
\left(-\frac{i}{2}\overleftrightarrow D\right)^2 t^a \chi \cdot
\chi^{\dag}\left(-\frac{i}{2}\overleftrightarrow D \times
\vec{\sigma}\right) t^a \psi + \textrm{H.c.}\,,
\\
\mathcal P_{8}(^3P_2)= & \frac{1}{2}
\psi^{\dag}\left(-\frac{i}{2}\overleftrightarrow D^{(i}
\sigma^{j)}\right) \left(- \frac{i}{2} \overleftrightarrow  D \right)^2
t^a \chi \,  \chi^{\dag}\left(-\frac{i}{2}\overleftrightarrow
D^{(i}\sigma^{j)}\right) t^a \psi + \textrm{H.c.}\,,
\\
\mathcal P_{8}(^3P_2,^3F_2)= & \frac{1}{2}
\psi^{\dag}\left(-\frac{i}{2}\right)^2\overleftrightarrow
D^{(i}\overleftrightarrow D^{j)}
\left(-\frac{i}{2}\overleftrightarrow D \cdot\vec{\sigma}\right)
t^a \chi \, \chi^{\dag}\left(-\frac{i}{2}\overleftrightarrow
D^{(i}\sigma^{j)}\right) t^a \psi
\\
& - \frac{1}{5}
\psi^{\dag}\left(-\frac{i}{2}\right)\overleftrightarrow D^{(i}
\sigma^{j)} \left(- \frac{i}{2} \overleftrightarrow  D \right)^2
t^a \chi \, \chi^{\dag}\left(-\frac{i}{2}\overleftrightarrow
D^{(i}\sigma^{j)}\right) t^a \psi + \textrm{H.c.}\,.
\end{split}
\ee
\be
\label{eq:B.6oct}
\begin{split}
\mathcal{Q}_{8}(^1D_2) = &
\psi^{\dag}\left(-\frac{i}{2}\right)^2\overleftrightarrow D^{(i}
\overleftrightarrow D^{j)} t^a \chi \,
\chi^{\dag}\left(-\frac{i}{2}\right)^2\overleftrightarrow D^{(i}
\overleftrightarrow D^{j)} t^a \psi,
\\
\mathcal Q_{8}(^3D_3)= & \psi^{\dag}\left(-\frac{i}{2}\right)^2
\overleftrightarrow D^{((i} \overleftrightarrow D^{j)}
\sigma^{l)} t^a \chi \, \chi^{\dag}\left(-\frac{i}{2}\right)^2
\overleftrightarrow D^{((i} \overleftrightarrow D^{j)}
\sigma^{l))} t^a \psi,
\\
\mathcal Q_{8}(^3D_2)= &
\frac{2}{3}
\psi^{\dag}
\left(-\frac{i}{2}\right)^2 \left(\varepsilon^{ilm}\overleftrightarrow D^{(j}  \overleftrightarrow D^{l)} \sigma^{m}
+ \frac{1}{2}\varepsilon^{ijl}\overleftrightarrow D^{(m}  \overleftrightarrow D^{l)} \sigma^{m}\right) t^a \chi
\\
& \times
 \chi^{\dag} \left(-\frac{i}{2}\right)^2 \left(\varepsilon^{inp}\overleftrightarrow D^{(j}  \overleftrightarrow D^{n)} \sigma^{p}
+ \frac{1}{2}\varepsilon^{ijn}\overleftrightarrow D^{(p}  \overleftrightarrow D^{n)} \sigma^{p}\right) t^a \psi,
\\
\mathcal Q_{8} (^3D_1) = & \psi^{\dag}\left(-\frac{i}{2}\right)^2
\overleftrightarrow D^{(i} \overleftrightarrow D^{j)}
\sigma^{i}\, t^a \chi \,  \chi^{\dag}\left(-\frac{i}{2}\right)^2
\overleftrightarrow D^{(l} \overleftrightarrow D^{j)}
\sigma^{l}\, t^a \psi.
\end{split}
\ee

\section{Summary of matching coefficients}
\label{AppB}
In the following, we list all the imaginary parts of the matching coefficients of the four-fermion operators
up to dimension 10, calculated at ${O}(\als^2)$ in the strong coupling constant
in section \ref{matching}.

In the presentation of the results we give for completeness also the matching coefficients obtained 
by using a basis of operators that includes
$1/2 \left(\mathcal Q'_{8}(^3S_1)\right.$ $\left.- \mathcal Q''_{8}(^3S_1) \right)$, 
$1/2 \left(\mathcal Q'_{8}(^3S_1,^3D_1)\right.$ $\left.- \mathcal Q''_{8}(^3S_1,^3D_1)\right)$ 
and $\mathcal T^{(1)\prime}_{8\textrm-1}(^3S_1,^3P)$ instead of $\mathcal T_{1\textrm-8}(^3P_0,^3S_1)$, 
$\mathcal T_{1\textrm-8}(^3P_1,$ $^3S_1)$, $\mathcal T_{1\textrm-8}(^3P_2,^3S_1)$. 
It is understood that when this basis is used, the coefficients 
$\textrm{Im}\, t_{1\textrm{-}8}(^3P_J,$ $^3S_1)$, 
with $J=0,1,2$ are set to 0. 
Viceversa if our basis contains the operators  $\mathcal T_{1\textrm-8}(^3P_J,^3S_1)$, 
with $J = 0,1,2$, the coefficients ${\rm Im} \,h'_{8}(^3S_1) - {\rm Im} \,h''_{8}(^3S_1)$, 
${\rm Im} \,h'_{8}(^3S_1,^3D_1) - {\rm Im} \,h''_{8}(^3S_1,^3D_1)$ 
and ${\rm Im} \,t^{(1)\prime}_{8\textrm-1}(^3S_1,^3P)$ are set to 0.

\begin{flalign*}
& \text{\bf Operator of dim. 6} &  & \textrm{\bf{Matching coefficient}} & \bf{Im\,(Value)}&                  & \\
&  \mathcal O_{1}(^1S_0)          &  & {\rm Im} \,f_{1}(^1S_0)            
& \als^2 \pi \, \frac{C_F}{2N_c} & \quad  \textrm{\cite{Bodwin:1994jh}} &	 \\
&  \mathcal O_{1}(^3S_1)          &  & {\rm Im} \,f_{1}(^3S_1)            &     0\quad &   
\end{flalign*}
\begin{flalign*}
& \hspace{3.7cm}               &  & \hspace{3.90cm}        &
\hspace{2cm}&  & \\
&  \mathcal O_{8}(^1S_0)       &  & {\rm Im} \,f_{8}(^1S_0)&
             \als^2 \pi  \, \frac{N^2_c-4}{4 N_c}&  \quad \textrm{\cite{Bodwin:1994jh}}&	 \\
&  \mathcal O_{8}(^3S_1)       &  & {\rm Im} \,f_{8}(^3S_1)
&             \frac{1}{6} \als^2 \pi \, n_f& \quad \textrm{\cite{Bodwin:1994jh}}&	 \\
\end{flalign*}
\begin{flalign*}
& \textrm{\bf Operator of dim. 8}                  &  &  \textrm{\bf{Matching coefficient}} & \bf{Im\,(Value)}&               & \\
&  \mathcal P_{1}(^1S_0)       &  & {\rm Im} \,g_{1}(^1S_0)
&               - \frac{4}{3}\als^2 \pi \, \frac{C_F}{2N_c}&  \quad \textrm{\cite{Bodwin:1994jh}}&	 \\
&  \mathcal P_{1}(^3S_1)       &  & {\rm Im} \,g_{1}(^3S_1)&               0\quad&\\
&  \mathcal P_{1}(^3S_1,^3D_1) &  & {\rm Im}  \,g_{1}(^3S_1,^3D_1)&        0\quad&\\ 
& \\
&  \mathcal O_{1}(^1P_1)       &  & {\rm Im} \,f_{1}(^1P_1) &              0\quad&\\
&  \mathcal O_{1}(^3P_0)       &  & {\rm Im} \,f_{1}(^3P_0) 
&              3 \als^2 \pi \,\frac{C_F}{2N_c}& \quad \textrm{\cite{Bodwin:1994jh}}&	 \\
&  \mathcal O_{1}(^3P_1)       &  & {\rm Im} \,f_{1}(^3P_1) &              0\quad&\\
&  \mathcal O_{1}(^3P_2)       &  & {\rm Im} \,f_{1}(^3P_2) 
&               \frac{4}{5} \als^2 \pi \, \frac{C_F}{2N_c}& \quad \textrm{\cite{Bodwin:1994jh}}&	  
\end{flalign*}
\begin{flalign*}
& \hspace{3.6cm}               &  & \hspace{3.90cm}        & \hspace{2cm}& &\\
&  \mathcal S_{1\textrm-8}(^1S_0,^3S_1)      &  
& {\rm Im} \,s_{1\textrm-8}(^1S_0,^3S_1) &        \frac{\als^2 \pi}{4N_c} \left( \frac{1}{3}n_f - N_c\right)&\\
&  \mathcal S_{1\textrm-8}(^3S_1,^1S_0)      &  
& {\rm Im} \,s_{1\textrm-8}(^3S_1,^1S_0) &       0 \quad &
\end{flalign*}
\begin{flalign*}
& \hspace{4.1cm}               &  & \hspace{3.90cm}        & \hspace{2cm}& &\\
&  \mathcal P_{8}(^1S_0)       &  & {\rm Im} \,g_{8}(^1S_0) 
&              - \frac{4}{3}\als^2 \pi \, \frac{N^2_c-4}{4N_c}& \quad \textrm{\cite{Bodwin:1994jh, Petrelli:1997ge}}&	 \\
&  \mathcal P_{8}(^3S_1)       &  & {\rm Im} \,g_{8}(^3S_1) 
&              - \frac{2}{9}\als^2 \pi \, n_f & \quad \textrm{\cite{Bodwin:1994jh, Petrelli:1997ge}}&\\
&  \mathcal P_{8}(^3S_1,^3D_1) &  & {\rm Im}  \,g_{8}(^3S_1,^3D_1)
&        - \frac{1}{6}\als^2 \pi \, n_f & \quad \textrm{\cite{Bodwin:1994jh, Petrelli:1997ge}}&
\end{flalign*}
\begin{flalign*}
& \hspace{3.6cm}               &  & \hspace{3.90cm}        & \hspace{2cm}& &\\
&  \mathcal O_{8}(^1P_1)       &  & {\rm Im} \,f_{8}(^1P_1) 
&                \frac{\als^2 \pi N_c}{12}& \quad \textrm{\cite{Bodwin:1994jh, Petrelli:1997ge}}&\\
&  \mathcal O_{8}(^3P_0)       &  & {\rm Im} \,f_{8}(^3P_0) 
&              3 \als^2 \pi \,\frac{N^2_c - 4}{4N_c}& \quad \textrm{\cite{Bodwin:1994jh, Petrelli:1997ge}}&\\
&  \mathcal O_{8}(^3P_1)       &  & {\rm Im} \,f_{8}(^3P_1) 
&              0\quad& \quad \textrm{\cite{Bodwin:1994jh, Petrelli:1997ge}}&\\
&  \mathcal O_{8}(^3P_2)       &  & {\rm Im} \,f_{8}(^3P_2) 
&               \frac{4}{5} \als^2 \pi \, \frac{N^2_c-4}{4N_c}& \quad \textrm{\cite{Bodwin:1994jh, Petrelli:1997ge}}&
\end{flalign*}
\begin{flalign*}
& \hspace{3cm}               &  & \hspace{4.50cm}        & \hspace{2cm}& & \\
&  \mathcal O_{1 \,\textrm{cm}}   &  & {\rm Im} \,f_{1 \,\textrm{cm}}  &         \frac{1}{4}\als^2 \pi \, \frac{C_F}{2N_c}& \\
&  \mathcal O'_{1 \,\textrm{cm}}  &  & {\rm Im} \,f'_{1 \,\textrm{cm}} &              0\quad&\\
&  \mathcal O_{8 \,\textrm{cm}}   &  & {\rm Im}  \,f_{8 \,\textrm{cm}} &         \frac{1}{4}\als^2 \pi \, \frac{N^2_c - 4}{4N_c}&\\
&  \mathcal O'_{8 \,\textrm{cm}}  &  & {\rm Im}  \,f'_{8 \,\textrm{cm}}&         \frac{1}{24}\als^2 \pi \, n_f &
\end{flalign*}
\begin{flalign*}
& \hspace{3cm}               &  & \hspace{4.50cm}        & \hspace{2cm}& &\\
&  \mathcal P_{1 a\,\textrm{cm}}       &  & {\rm Im} \,g_{1a \,\textrm{cm}} &                0\quad&\\
&  \mathcal P_{1 b\,\textrm{cm}}       &  & {\rm Im} \,g_{1b \,\textrm{cm}} &                0\quad&\\
&  \mathcal P_{1 c\,\textrm{cm}}       &  & {\rm Im} \,g_{1c \,\textrm{cm}} &                -\frac{1}{4} \als^2 \pi \,\frac{C_F}{2N_c}&\\
&  \mathcal P_{8 a\,\textrm{cm}}       &  & {\rm Im} \,g_{8a \,\textrm{cm}} &                -\frac{1}{24} \als^2 \pi \,n_f& \\
&  \mathcal P_{8 b\,\textrm{cm}}       &  & {\rm Im} \,g_{8b \,\textrm{cm}} &               0\quad& \\
&  \mathcal P_{8 c\,\textrm{cm}}       &  & {\rm Im} \,g_{8c \,\textrm{cm}}&               -\frac{1}{4}\als^2 \pi \, \frac{N^2_c-4}{4N_c}& \\
\end{flalign*}

\vfill\eject

\begin{flalign*}
& \textrm{\bf Operator of dim. 9}  \qquad                 &  &  \textrm{\bf{Matching coefficient}} & \bf{Im\,(Value)}&   &\\
& \mathcal T^{(1) \prime}_{1\textrm-8}(^3S_1,^3P) &  & {\rm Im} \,t^{(1) \prime}_{1\textrm-8}(^3S_1,^3P)                     
& - \frac{1}{8} \als^2 \pi \frac{N^2_c - 4}{4N_c^2} & \\
& \mathcal T^{(1) \prime}_{8\textrm-1}(^3S_1,^3P) &  & {\rm Im} \,t^{(1) \prime}_{8\textrm-1}(^3S_1,^3P)                     
&  \frac{1}{24} \als^2 \pi \frac{n_f}{N_c} + \frac{1}{48} \als^2 \pi - \frac{1}{8} \als^2 \pi \frac{C_F}{N_c}&  \\
&  \mathcal T_{1\textrm-8}(^3P_0,^3S_1)      &  & {\rm Im} \,t_{1\textrm-8}(^3P_0,^3S_1)        
& - \frac{3}{2} \als^2 \pi \frac{C_F}{2N_c} + \left( \frac{61}{240} + \frac{7}{192} \frac{n_f}{N_c}\right) \als^2 \pi& \\
&  \mathcal T_{1\textrm-8}(^3P_1,^3S_1)      &  & {\rm Im} \,t_{1\textrm-8}(^3P_1,^3S_1)        
&  \left( \frac{1}{72} + \frac{107}{576} \frac{n_f}{N_c}\right) \als^2 \pi &\\
&    \mathcal T_{1\textrm-8}(^3P_2,^3S_1)    &  & {\rm Im} \,t_{1\textrm-8}(^3P_2,^3S_1)        
&  \left( \frac{1}{10} + \frac{25}{576} \frac{n_f}{N_c}\right) \als^2 \pi &\\
\end{flalign*}
\begin{flalign*}
& \textrm{\bf Operator of dim. 10}  \quad \, \,                &  &  \textrm{\bf{Matching coefficient}} \qquad \qquad& \bf{Im\,(Value)}&   & \\
&\mathcal Q'_{1}(^1S_0)        &  & {\rm Im} \,h'_{1}(^1S_0)            
&  \frac{10}{9}\als^2 \pi \frac{C_F}{2N_c} - \frac{1}{48}\als^2 \pi&  \quad \textrm{\cite{Bodwin:2002hg}}&\\
&\mathcal Q''_{1}(^1S_0)       &  & {\rm Im} \,h''_{1}(^1S_0)           
&  \frac{2}{5}\als^2 \pi \frac{C_F}{2N_c} + \frac{1}{48}\als^2 \pi & \quad \textrm{\cite{Bodwin:2002hg}}&\\
&\mathcal Q'_{1}(^3S_1)        &  & {\rm Im} \,h'_{1}(^3S_1)            
&  \frac{1}{12} \als^2 \pi \, \frac{N^2_c - 4}{4N_c^2}& \quad \textrm{\cite{Bodwin:2002hg}}&\\
& \mathcal Q''_{1}(^3S_1)      &  & {\rm Im} \,h''_{1}(^3S_1)           
& -\frac{1}{12} \als^2 \pi \,\frac{N^2_c - 4}{4N_c^2}& \quad \textrm{\cite{Bodwin:2002hg}}&\\
&  \mathcal Q'_{1}(^3S_1,^3D_1)&  & {\rm Im} \,h'_{1}(^3S_1,^3D_1)      
&  \frac{1}{4} \als^2 \pi \, \frac{N^2_c - 4}{4N_c^2}& \quad \textrm{\cite{Bodwin:2002hg}}&\\
& \mathcal Q''_{1}(^3S_1,^3D_1)&  & {\rm Im} \,h''_{1}(^3S_1,^3D_1)     
& -\frac{1}{4} \als^2 \pi \, \frac{N^2_c - 4}{4N_c^2}& \quad \textrm{\cite{Bodwin:2002hg}}&
\end{flalign*}
\begin{flalign*}
& \hspace{3.5cm}               &  & \hspace{5.5	cm}        &
\hspace{2cm}&  &\\
&\mathcal P_{1}(^1P_1)         &  & {\rm Im} \,g_{1}(^1P_1)             & 0\quad& \\
&\mathcal P_{1}(^3P_0)         &  & {\rm Im} \,g_{1}(^3P_0)             & - 7 \als^2 \pi \, \frac{C_F}{2N_c}&\\
& \mathcal P_{1}(^3P_1)        &  & {\rm Im} \,g_{1}(^3P_1)             & 0\quad&\\
&  \mathcal P_{1}(^3P_2)       &  & {\rm Im} \,g_{1}(^3P_2)             & -\frac{8}{5}\als^2 \pi \, \frac{C_F}{2N_c}& \\
&  \mathcal P_{1}(^3P_2,^3F_2) &  & {\rm Im} \,g_{1}(^3P_2,^3F_2)&  -\frac{20}{21}\als^2 \pi \, \frac{C_F}{2N_c} & 
\end{flalign*}
\begin{flalign*}
& \hspace{4.5cm}               &  & \hspace{6cm}        & \hspace{2cm}&  \hspace{2cm} &\\
&\mathcal Q_{1}(^1D_2)  \quad   \qquad     &  & {\rm Im} \,h_{1}(^1D_2)             
& \qquad \frac{2}{15} \als^2 \pi \,\frac{C_F}{2N_c}& \quad \textrm{\cite{Novikov:1977dq}}&\\
&\mathcal Q_{1}(^3D_1)         &  & {\rm Im} \,h_{1}(^3D_1)             & 0\quad&\\
& \mathcal Q_{1}(^3D_2)        &  & {\rm Im} \,h_{1}(^3D_2)             & 0\quad&\\
&  \mathcal Q_{1}(^3D_3)       &  & {\rm Im} \,h_{1}(^3D_3)             & 0\quad&
\end{flalign*}

\begin{flalign*}
& \hspace{3.7cm}               &  & \hspace{3.90cm}                     &   \hspace{2cm}  &\\
&\mathcal Q'_{8}(^1S_0)        &  & {\rm Im} \,h'_{8}(^1S_0)            &  \hspace{-1cm} \frac{10}{9}\als^2 \pi \frac{N^2_c-4}{4N_c} &\\
&\mathcal Q''_{8}(^1S_0)       &  & {\rm Im} \,h''_{8}(^1S_0)           &   \hspace{-1cm} \frac{2}{5}\als^2 \pi \frac{N^2_c-4}{4N_c}  &\\
&\frac{  \mathcal Q'_{8}(^3S_1) +  \mathcal Q''_{8}(^3S_1) }{2}   &  & {\rm Im} \,h'_{8}(^3S_1) +  {\rm Im} \,h''_{8}(^3S_1)            
&   \hspace{-1cm} \frac{29}{108} \als^2 \pi n_f + \frac{1}{108} \als^2 \pi N_c & \\
&  \frac{\mathcal Q'_{8}(^3S_1,^3D_1) + \mathcal Q''_{8}(^3S_1,^3D_1)}{2}&  & {\rm Im} \,h'_{8}(^3S_1,^3D_1) +  {\rm Im} \,h''_{8}(^3S_1,^3D_1)      
&   \hspace{-1cm} \frac{23}{72} \als^2 \pi n_f + \frac{1}{18} \als^2 \pi N_c &\\
& \frac{\mathcal Q'_{8}(^3S_1) - \mathcal Q''_{8}(^3S_1)   }{2}     &  & {\rm Im} \,h'_{8}(^3S_1) - {\rm Im} \,h''_{8}(^3S_1)            
& \hspace{2cm}& \\
&\hspace{3.7cm}               &  & \hspace{3.90cm}                  
&   \hspace{-2cm} \frac{17}{108} \als^2 \pi n_f - \frac{41}{108} \als^2 \pi N_c +  \frac{1}{3} \als^2 \pi C_F& \\
& \frac{ \mathcal Q'_{8}(^3S_1,^3D_1) - \mathcal Q''_{8}(^3S_1,^3D_1) }{2} &  & {\rm Im} \,h'_{8}(^3S_1,^3D_1) - {\rm Im} \,h''_{8}(^3S_1,^3D_1)     
& \hspace{2cm}& \\
&\hspace{3.7cm}               &  & \hspace{3.90cm}                  
& \hspace{-2cm} \frac{17}{72} \als^2 \pi n_f - \frac{23}{45}  \als^2 \pi N_c + \als^2 \pi C_F&
\end{flalign*}
\begin{flalign*}
& \hspace{3.7cm}               &  & \hspace{3.90cm}        & \hspace{2cm}&  &\\
&\mathcal P_{8}(^1P_1)         &  & {\rm Im} \,g_{8}(^1P_1)             & -\frac{3}{20} \als^2 \pi N_c& \\
&\mathcal P_{8}(^3P_0)         &  & {\rm Im} \,g_{8}(^3P_0)             & - 7 \als^2 \pi \, \frac{N^2_c - 4}{4N_c}&\\
& \mathcal P_{8}(^3P_1)        &  & {\rm Im} \,g_{8}(^3P_1)             & 0\quad&\\
&  \mathcal P_{8}(^3P_2)       &  & {\rm Im} \,g_{8}(^3P_2)             & -\frac{8}{5}\als^2 \pi \, \frac{N^2_c - 4}{4N_c}& \\
&  \mathcal P_{8}(^3P_2,^3F_2) &  & {\rm Im} \,g_{8}(^3P_2,^3F_2)       &-\frac{20}{21}\als^2 \pi \, \frac{N^2_c - 4}{4N_c} & 
\end{flalign*}
\begin{flalign*}
& \hspace{3.7cm}               &  & \hspace{3.90cm}        & \hspace{2cm}&  &\\
&\mathcal Q_{8}(^1D_2)         &  & {\rm Im} \,h_{8}(^1D_2)             & \frac{2}{15} \als^2 \pi \,\frac{N^2_c - 4}{4N_c}& \\
&\mathcal Q_{8}(^3D_1)         &  & {\rm Im} \,h_{8}(^3D_1)             & \frac{1}{24} \als^2 \pi n_f + \frac{1}{12} \als^2 \pi N_c&\\
& \mathcal Q_{8}(^3D_2)        &  & {\rm Im} \,h_{8}(^3D_2)             & \frac{1}{30} \als^2 \pi N_c &\\
&  \mathcal Q_{8}(^3D_3)       &  & {\rm Im} \,h_{8}(^3D_3)             & \frac{1}{21} \als^2 \pi N_c&
\end{flalign*}

\end{document}